\documentclass[sigconf]{acmart} 

\usepackage{booktabs}   
\usepackage{subcaption} 

\usepackage{enumitem}
\setlist{leftmargin=5.5mm}

\usepackage{fancyvrb}
\usepackage{ifthen}
\usepackage{wrapfig}

\newcommand{\snsScreenshotScale}{0.35}
\newcommand{\lambdaScreenshotScale}{0.265}
\newcommand{\lambdaScreenshotScaleBigger}{0.30}
\newcommand{\numUsers}{21}
\newcommand{\numExamples}{five}

\newcommand{\exampleZero}{\sns{} Logo}
\newcommand{\exampleOne}{Target}
\newcommand{\exampleTwo}{Battery Icon}
\newcommand{\exampleThree}{Coffee Mug}
\newcommand{\exampleFour}{Mondrian Arch}
\newcommand{\locExamples}{100} 
\newcommand{\timeRawVideos}{35}
\newcommand{\timeRawVideoBattery}{15}

\newcommand{\locImplementation}{9,000}

\newcommand{\maybeCode}{Code}
\newcommand{\maybecode}{code}

\newcommand{\suppMaterials}
  {Supplementary Appendices~\cite{sns-deuce-full}}

\def\parahead#1{\paragraph{\textbf{#1.}}}
\def\paraheadNoDot#1{\paragraph{\textbf{#1}}}

\newcommand{\ie}{{\emph{i.e.}}}
\newcommand{\eg}{{\emph{e.g.}}}

\newcommand{\cf}{{\emph{cf.}}}

\newcommand{\little}{\mbox{\textsc{Little}}}
\newcommand{\sns}{\ensuremath{\textsc{Sketch-n-Sketch}}}
\newcommand{\version}[1]{\ensuremath{\textsc{v#1}}}
\newcommand{\deuce}{\ensuremath{\textsc{Deuce}}}
\newcommand{\dndr}{\ensuremath{\textsc{DNDRefactoring}}}
\newcommand{\hare}{\ensuremath{\textsc{HaRe}}}

\newcommand{\codeTool}[1]
  {\textsl{#1}}

\newcommand{\sep}{\hspace{0.06in}}

\newcommand{\vsepRuleHeight}{0.12in}
\newcommand{\vsepRule}{\vspace{\vsepRuleHeight}}
\newcommand{\miniSepOne}{\hspace{0.01in}}
\newcommand{\miniSepTwo}{\hspace{0.02in}}
\newcommand{\miniSepThree}{\hspace{0.03in}}

\newcommand{\maybeOptions}[1]{\ifthenelse{\equal{#1}{}}{}{ (+ {#1})}}
\newcommand{\maybeSelectedEquations}[1]{\ifthenelse{\equal{#1}{0}}{}{ (or {#1})}}

\newcommand{\beginToolTable}{\begin{tabular}{p{3.5in}|ccc}}
\newcommand{\rowToolNoBreak}[6]{\codeTool{#1}\maybeOptions{#6}&{#2}&{#3}\maybeSelectedEquations{#4}&{#5}}
\newcommand{\rowTool}[6]{\rowToolNoBreak{\codeTool{#1}}{#2}{#3}{#4}{#5}{#6}\\}
\newcommand{\header}[1]{\textbf{#1}}
\newcommand{\toolTableHeaders}
  {\multicolumn{1}{l|}{\textit{\makebox{Active Transformations}}} &
   \multicolumn{3}{l}{\textit{Selected Widgets}} \\
   \rowTool{\header{Tool Name}}
           {\header{Expressions}}
           {\header{Patterns}}
           {\header{Defs}}
           {\header{Targets}}
           {\header{Options}}\hline}
\newcommand{\toolCategory}[1]{\multicolumn{4}{l}{#1} \\ \hline}
\newcommand{\toolCategoryReuse}
  {\toolCategory{Preserve Behavior for Reuse}}
\newcommand{\toolCategoryStyle}
  {\toolCategory{Preserve Behavior for Readability or Subsequent Editing}}
\newcommand{\toolCategoryChange}
  {\toolCategory{Change Behavior}}

\newcommand{\varConst}{c}

\newcommand{\varExp}{\textbf{e}}
\newcommand{\varExpBare}{e}

\newcommand{\varVar}{x}
\newcommand{\varPat}{\textbf{p}}
\newcommand{\varPatBare}{p}

\newcommand{\varExpId}{i}

\newcommand{\varProgram}{\mathit{program}}

\newcommand{\ttlparen}{\ensuremath{\texttt{(}}}
\newcommand{\ttrparen}{\ensuremath{\texttt{)}}}
\newcommand{\ttlbrack}{\ensuremath{\texttt{[}}}
\newcommand{\ttrbrack}{\ensuremath{\texttt{]}}}
\newcommand{\ttpipe}{\ensuremath{\texttt{|}}}

\newcommand{\targetPosBefore}{\ensuremath{\textcolor{cyan}{\bullet}}}
\newcommand{\targetPosAfter}{\ensuremath{\targetPosBefore}}
\newcommand{\targetsAround}[1]
  {\targetPosBefore\miniSepTwo{#1}\miniSepTwo\targetPosAfter}
\newcommand{\targetBefore}[1]
  {\targetPosBefore\miniSepTwo{#1}}

\newcommand{\parens}[1]{\ensuremath{\ttlparen{#1}\ttrparen}}
\newcommand{\bracks}[1]{\ensuremath{\ttlbrack{#1}\ttrbrack}}
\newcommand{\spaceOne}[1]{\ensuremath{\miniSepOne{#1}\miniSepOne}}
\newcommand{\expFun}[2]{\ensuremath{\expTwoThings{\lambda}{#1}{#2}}}
\newcommand{\expNil}{\ensuremath{\ttlbrack\ttrbrack}}
\newcommand{\expCons}[2]{\ensuremath{\bracks{\spaceOne{#1}\ttpipe\spaceOne{#2}}}}

\newcommand{\expListTwo}[2]{\bracks{\spaceOne{\twoThings{#1}{#2}}}}
\newcommand{\expApp}[2]{\ensuremath{\parens{{#1}\miniSepThree{ }{#2}}}}

\newcommand{\expStr}[1]
  {\ensuremath{\texttt{'{#1}'}}}

\newcommand{\wrapInBox}[1]{\fcolorbox{orange}{white}{#1}}

\newcommand{\expLetNoBox}[3]
  {\parens{\twoThings{{\texttt{let}\miniSepThree\ensuremath{{#1}\miniSepThree{#2}}}}{#3}}}
\newcommand{\expLet}[3]
  {\parens{\twoThings{\wrapInBox{\texttt{let}\miniSepThree\ensuremath{{#1}\miniSepThree{#2}}}}{#3}}}

\newcommand{\expTopDefNoBox}[2]{\expTwoThings{\texttt{def}}{#1}{#2}}
\newcommand{\expTopDef}[2]{\wrapInBox{\expTwoThings{\texttt{def}}{#1}{#2}}}
\newcommand{\expDef}[3]{\twoThings{\expTopDef{#1}{#2}}{#3}}

\newcommand{\expCaseTwo}[3]{\expThreeThings{\texttt{case}}{#1}{#2}{#3}}

\newcommand{\expBranchWithId}[4]{\wrapInBox{\expOneThing{#3}{#4}}}

\newcommand{\twoThings}[2]{\ensuremath{{#1}\miniSepThree{#2}}}
\newcommand{\threeThings}[3]{\ensuremath{{#1}\miniSepThree\twoThings{#2}{#3}}}
\newcommand{\fourThings}[4]{\ensuremath{{#1}\miniSepThree\threeThings{#2}{#3}{#4}}}

\newcommand{\expOneThing}[2]{\ensuremath{\parens{\twoThings{#1}{#2}}}}
\newcommand{\expTwoThings}[3]{\ensuremath{\parens{\threeThings{#1}{#2}{#3}}}}
\newcommand{\expThreeThings}[4]{\ensuremath{\parens{\fourThings{#1}{#2}{#3}{#4}}}}

\newcommand{\figSyntaxLineBreak}{\\[1pt]}
\newcommand{\figSyntaxSpaceNextCategory}{\\[1pt]}

\newcommand{\figSyntaxSpaceItem}{\sep\mid\sep}

\newcommand{\figSyntaxEnd}{\end{array}$}

\newcommand{\figSyntaxBegin}{$\begin{array}{rcl}}
\newcommand{\figSyntaxRowLabel}[2]{{#2}&\miniSepTwo{::=}\miniSepTwo&}
\newcommand{\figSyntaxRow}{&\miniSepOne\mid\miniSepOne&}

\newcommand{\aSubst}[2]{{#1}\mapsto{#2}}

\newcommand{\helperop}[1]{\ensuremath{\mathsf{#1}}}

\newcommand{\myEquation}[2]{\ensuremath{{#1}={#2}}}

\newcommand{\solveOne}[4]{\ensuremath{\helperop{SolveOne}(#1,#2,\myEquation{#3}{#4})}}

\newcommand{\solveAndUpdate}[4]
  {(\aSubst{#2}{\solveOne{#1}{#2}{#3}{#4}})}

\copyrightyear{2018}
\acmYear{2018}
\setcopyright{acmlicensed}
\acmConference[ICSE '18]
  {ICSE '18: 40th International Conference on Software Engineering }
  {May 27--June 3, 2018}{Gothenburg, Sweden}
\acmBooktitle
  {ICSE '18: 40th International Conference on Software
   Engineering , May 27--June 3, 2018, Gothenburg, Sweden}
\acmPrice{15.00}
\acmDOI{10.1145/3180155.3180165}
\acmISBN{978-1-4503-5638-1/18/05}

\begin{document}

\title{\deuce{}: A Lightweight User Interface for Structured Editing}

\author{Brian Hempel, Justin Lubin, Grace Lu, and Ravi Chugh}
\affiliation{\institution{University of Chicago}}
\email{{brianhempel,justinlubin,gracelu,rchugh}@uchicago.edu}

\begin{abstract}

We present a structure-aware code editor, called \deuce{}, that is
equipped with direct manipulation capabilities for invoking
automated program transformations. Compared to traditional
refactoring environments, \deuce{} employs a direct manipulation
interface that is tightly integrated within a text-based editing
workflow. In particular, \deuce{} draws (i) clickable widgets atop the
source code that allow the user to \emph{structurally select} the
unstructured text for subexpressions and other relevant features, and
(ii) a lightweight, interactive menu of potential transformations
based on the current selections.
We implement and evaluate our design
with mostly standard transformations
in the context of a small functional programming language.
A controlled user study with \numUsers{} participants demonstrates
that structural selection is preferred to a more traditional
text-selection interface and may be faster overall once users
gain experience with the tool.
These results accord with \deuce{}'s aim to provide human-friendly
structural interactions on top of familiar text-based editing.





\end{abstract}

\ccsdesc{Software and its engineering~Integrated and visual development environments}
\ccsdesc{Human-centered computing~Human computer interaction (HCI)}

\keywords{Structured Editing, Refactoring, Direct Manipulation}

\maketitle

\section{Introduction}
\label{sec:intro}

Plain text continues to dominate as the universal format for programs
in most languages. Although the simplicity and generality of text are
extremely useful, the benefits come at some costs. For novice
programmers, the unrestricted nature of text leaves room for syntax errors
that make learning how to program more difficult~\cite{GreenfootCost}.
For expert programmers, many editing tasks---perhaps even the vast
majority~\cite{Ko2005}---fall within specific patterns that could be
performed more easily and safely by automated tools. Broadly speaking,
two lines of work have, respectively, sought to address these
limitations.

\paragraph{Structured Editing}

Structured editors---such as
the Cornell Program Synthesizer~\cite{Teitelbaum1981},
Scratch~\cite{Scratch:2009,Scratch:2010}, and
TouchDevelop~\cite{TouchDevelop}---reduce the
amount of unstructured text used to represent programs, relying on
blocks and other visual elements to demarcate structural components of
a program (\eg{} a conditional with two branches, and a function with an
argument and a body). Operations that create and manipulate structural
components avoid classes of errors that may otherwise arise in plain
text, and text-editing is limited to within well-formed structures.
Structured editing has not yet, however, become popular among expert
programmers, in part due to their cumbersome interfaces compared to
plain text editors~\citep{Monig:2015}, as well as their restrictions
that even transitory, evolving programs always be well-formed.

\paragraph{Text Selection-Based Refactoring}

An alternative approach in integrated development environments (IDEs),
such as Eclipse, is to augment unrestricted plain text with support
for a variety of
refactorings~\cite{GriswoldThesis,Fowler1999,SmalltalkRefactoring}. In such
systems, the user text-selects something of interest in the
program---an expression, statement, type, or class---and then selects
a transformation either from a menu at the top of the IDE or
in a right-click pop-up menu. This approach provides experts both the full
flexibility of text as well as mechanisms to perform common tasks more
efficiently and with fewer errors than with manual, low-level text-edits.
Although useful, this workflow suffers limitations:

\begin{enumerate}

\item The text-selection mechanism is error-prone when the item to select is
long, spanning a non-rectangular region or requiring
scrolling~\citep{Murphy-Hill-ICSE2008}.

\item All transformations must require a single ``primary'' selection argument,
and any additional arguments are relegated to a separate Configuration Wizard
window.

\item The list of tools is typically very long---even in the right-click menu
where tools that are not applicable to the primary selection are filtered
out---making it hard to \emph{identify}, \emph{invoke}, and \emph{configure} a
desired refactoring~\citep{Murphy-Hill:2009,Vakilian:2012,Mealy:2007}.

\item Even when a transformation has no configuration options or when the
defaults are acceptable---as is often the case~\citep{Murphy-Hill:2009}---the
user must go through a separate Configuration Wizard to make the change. The
user must, furthermore, navigate to another pane within the Configuration Wizard
to preview the changes before confirming them.

\end{enumerate}

\parahead{Our Approach}

Our goal is to enable a workflow that enjoys the benefits
of both approaches. Specifically, programs ought to be represented in
plain text for familiar and flexible editing by expert programmers,
and the editing environment ought to provide automated support for a
variety of code transformations without deviating
from the text-editing workflow.

In this paper, we present a structure-aware editor, called \deuce{}, that
achieves these goals by augmenting a text editor with (i) clickable
widgets directly atop the program text that allow the user to \emph{structurally
select} the unstructured text for subexpressions and other relevant features of
the program structure, and (ii) a \emph{context-sensitive tool menu with
previews} based on the current selections.

\paragraph{Structural \maybeCode{} Selection}

Rather than relying on keyboard-based text-edits for selection, our editor draws
direct manipulation widgets to \emph{structurally select} items in the code with
a single mouse-click. In particular, holding down the Shift key transitions the
editor into structural selection mode. In this mode, the editor draws a box
(which resembles a text-selection highlight) around the code item below the
current mouse position. Clicking the box selects the entire text for that code
item, eliminating any possibility for error and reducing the time needed to
select long, non-rectangular sequences of lines. Furthermore, this interface naturally
allows multiple selection, even when items are far apart in the code. Structural
text selection helps address concerns (1) and (2) above.

\paragraph{Context-Sensitive Menu with Previews}

Because structural selection naturally supports multiple selection, we
address concern (3) by showing only tools for which \emph{all} necessary
arguments have been selected, reducing the number of tools shown to the user
compared to a typical right-click menu.
Hovering over a result description previews the changes, and
clicking a result chooses it. For tools with few
configuration options, we believe the preview menu provides a lightweight way to
consider multiple options while staying within the normal editing workflow,
helping to address concern (4).
\\

\noindent The resulting workflow in \deuce{} is largely
text-driven, but augmented with automated support for code transformations
(\eg{} to introduce local variables, rearrange definitions, and introduce
function abstractions) that are tedious and error-prone (\eg{} because of typos,
name collisions, and mismatched delimiters), allowing the user to spend
keystrokes on more creative and difficult tasks that are harder to automate.
The name \deuce{} reflects this streamlined combination of text- and mouse-based
editing.

\parahead{Contributions}

This paper makes the following contributions:

\begin{itemize}

\item We present the design of \deuce{}, a code editor equipped with
\emph{structural \maybecode{} selection}, a lightweight direct manipulation mechanism
that helps to identify and invoke program transformations while
retaining the freedom and familiarity of traditional text-based editing. Our
design can be instantiated for different programming languages and with
different sets of program transformations.
(\autoref{sec:user-interface})

\item We implement \deuce{} within \sns{}, a programming
environment for creating Scalable Vector Graphics (SVG) images.
Most of the functional program transformations in our implementation
are common to existing refactoring tools, but two
transformations---\codeTool{Move Definitions} and \codeTool{Make Equal}---are,
to the best of our knowledge, novel.
(\autoref{sec:little-transformations})

\item To evaluate the utility of our user interface, we
performed a controlled user study with \numUsers{} participants. The results
show that, compared to a more traditional text selection-based refactoring
interface, structural \maybecode{} selection is
preferred and may be faster for invoking
transformations, particularly as users gain experience with the tool.
(\autoref{sec:user-study})

\end{itemize}

\noindent
Our implementation, videos of examples, and
user study materials are available at
\url{http://ravichugh.github.io/sketch-n-sketch/}.
In the next section, we introduce \deuce{} with a few short examples.

\section{Overview Examples}
\label{sec:overview}

\setlength{\intextsep}{6pt}%
\setlength{\columnsep}{10pt}%

\newcommand{\overviewExample}[2]{\paragraph{Example {#1}}}

\overviewExample{1}{Red Square}

\begin{wrapfigure}{r}{0pt}
\includegraphics[scale=\lambdaScreenshotScale]{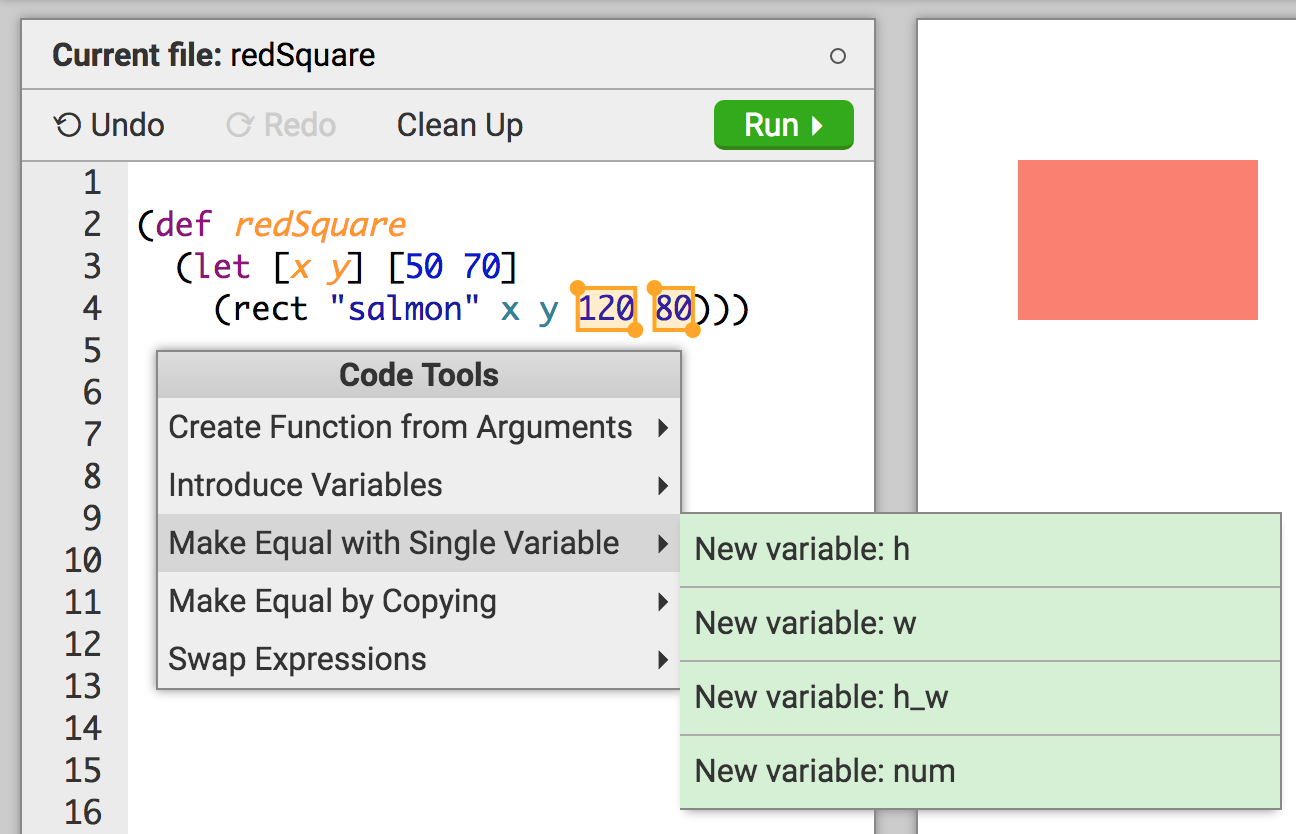}
\end{wrapfigure}
Despite the intention of the following program,
the \verb+redSquare+ definition uses different values
for the width and height of the rectangle
(the fourth and fifth arguments, respectively, to the
\verb+rect+ function).
The user chooses \deuce{} code tools---rather than
text-edits---to correct this mistake.

The user presses the Shift key to enter structured editing mode, and then hovers
over and clicks the two constants \verb+120+ and \verb+80+ to select them; the
selected code items are colored orange in the screenshot above.
Based on these selections, \deuce{} shows a pop-up Code Tools menu with several
potential transformations.
The \codeTool{Make Equal by Copying} tool would replace one of the constants with the
other, thus generating a square. However, such a program would require two
constants to be changed whenever a different size is desired.
Instead, the user wishes to invoke \codeTool{Make Equal with Single Variable} to introduce a
new variable that will be used for both arguments. Hovering over this menu item
displays a second-level menu (shown above) with tool-specific options, in this
case, the names of four suggested new variable names.

\begin{wrapfigure}{r}{0pt}
\includegraphics[scale=\lambdaScreenshotScaleBigger]{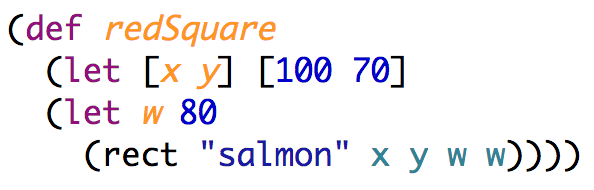}
\end{wrapfigure}
The user hovers over the second option, which shows a preview of
the transformed code (shown on the right).
The user clicks to choose the second option.
Notice that the number \verb+80+ (rather than \verb+120+) was
chosen to be the value of the new variable \verb+w+.
Whereas the tool provided configuration options for the
variable name, it did not provide options for which value to use;
this choice was made by the implementor of the \codeTool{Make Equal} code tool,
not by the \deuce{} user interface.

\overviewExample{2}{Two Circles}

Consider the following program that draws two circles connected by a line.
All design parameters and shapes have been organized within
a single top-level \verb+connectedCircles+ definition.
To make the design more reusable, the user wants
\verb+connectedCircles+ to be a function that is abstracted over the positions
of the two circles.
The user hovers over and clicks the \verb+def+ keyword, and selects the \codeTool{Create Function
from Definition} tool
(shown in the screenshot).

\begin{center}
\includegraphics[scale=\lambdaScreenshotScale]{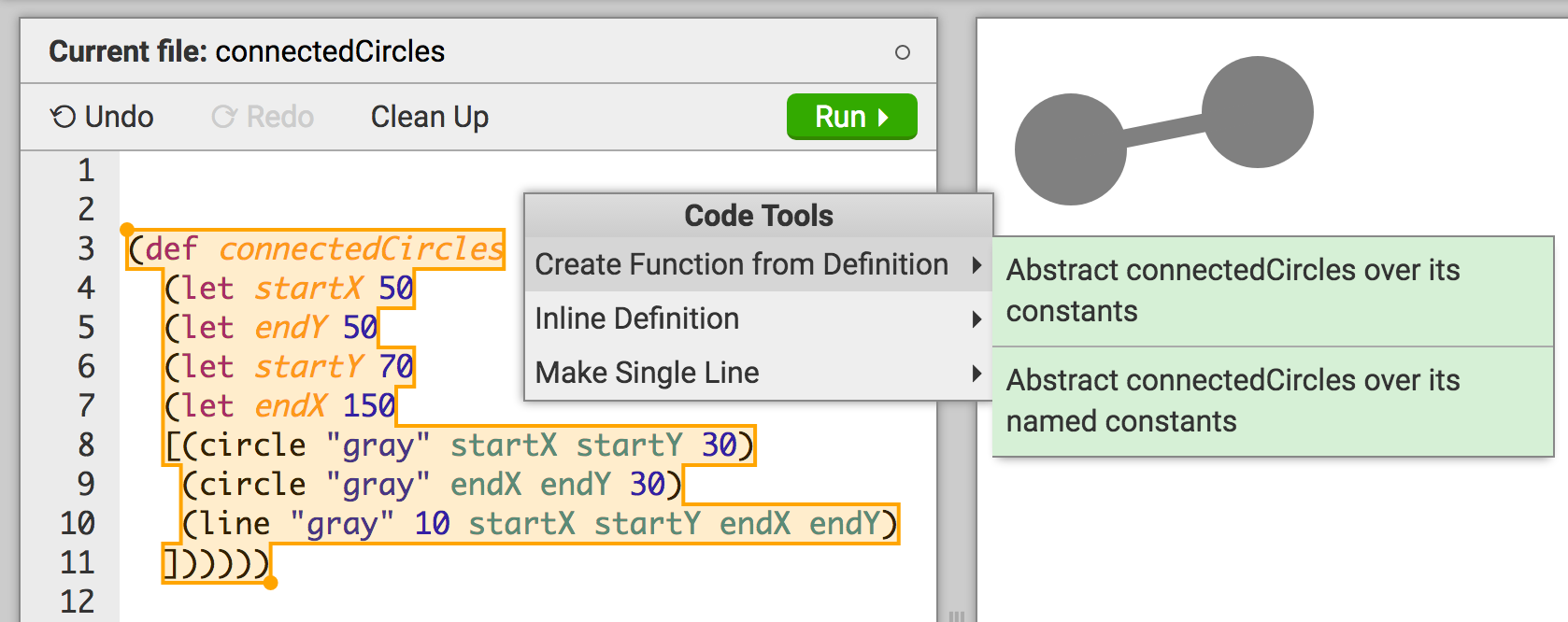}
\end{center}

\begin{wrapfigure}{r}{0pt}
\includegraphics[scale=\lambdaScreenshotScaleBigger]{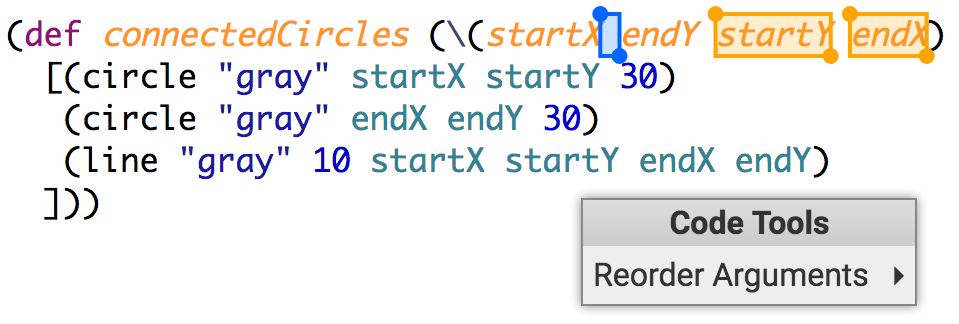}
\end{wrapfigure}
\noindent
In response, \codeTool{Create Function}
rewrites the definition
to be a function (shown on the right), and
previous uses of \verb+connectedCircles+ are rewritten to
appropriate function calls (not shown).

The order of arguments to the function match the order of definitions
in the previous program, but that order was unintuitive---the coordinates of the
\verb+start+ and \verb+end+ points were interleaved. To fix this, as shown above, the
user selects the last two arguments and the \emph{target position} (\ie{} the
space enclosed by a blue rectangular selection widget) between the first two, and
selects the \codeTool{Reorder Arguments} tool so that the order of arguments becomes
\verb+startX+, \verb+startY+, \verb+endX+, and \verb+endY+ (not shown).
Calls to \verb+connectedCircles+ are, again, rewritten to match the new order
(not shown).

\overviewExample{3}{Lambda Icon}

In the program below, the user would like to organize all design parameters and
shapes within the single \verb+logo+ definition. The user hovers over and
selects the five definitions on lines 2 through 9, as well as the space
on line 13, and selects the \codeTool{Move Definitions} tool to move the definitions inside
\verb+logo+. The transformation manipulates indentation and
delimiters appropriately in the final code (not shown).

\vspace{2pt} 

\begin{center}
\includegraphics[scale=\lambdaScreenshotScale]{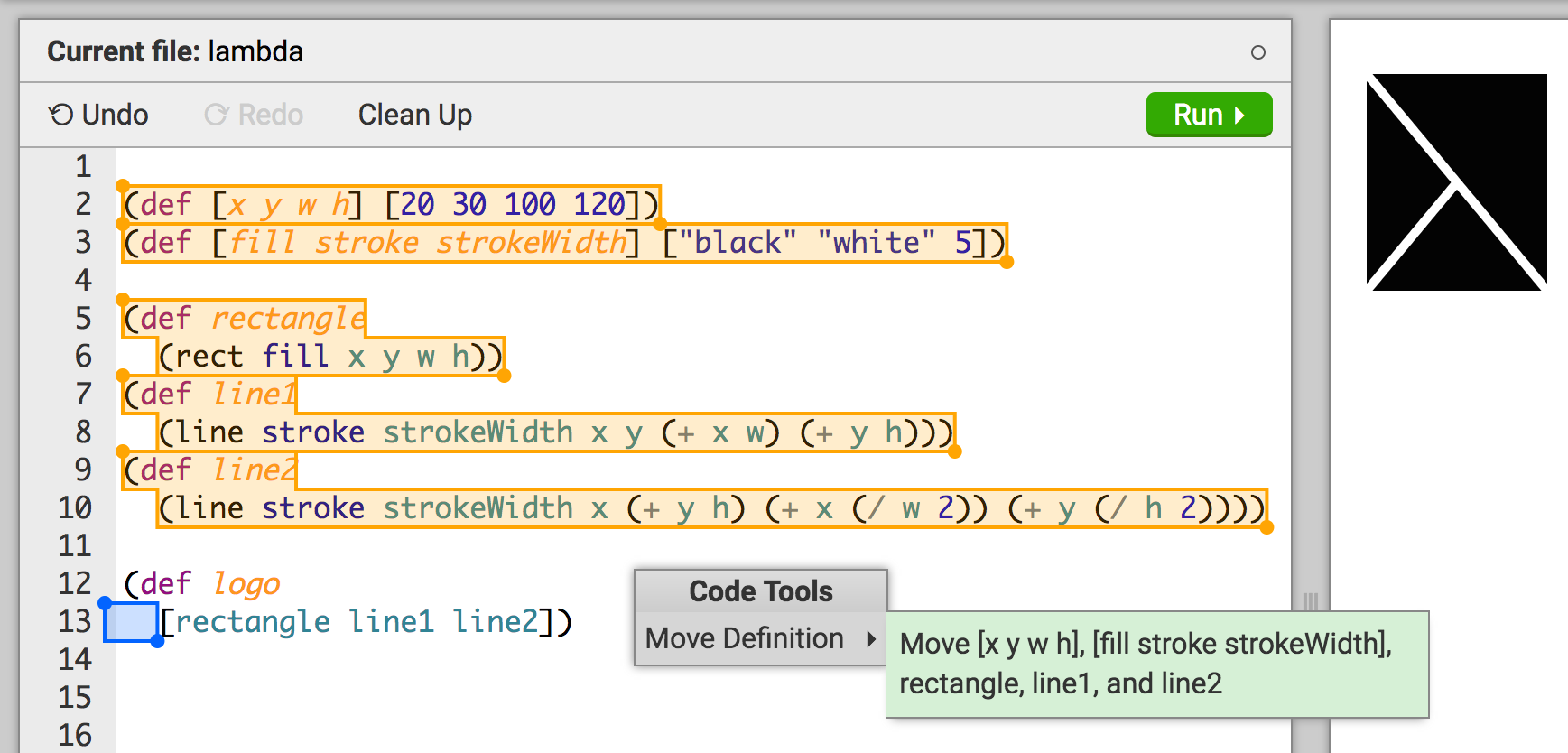}
\end{center}

\section{\deuce{}: Design and Implementation}
\label{sec:deuce}

In this section, we explain the design of \deuce{} in more detail.
First, we define a core language of programs where various structural
features can be selected.
Then, we describe a user interface that displays 
active transformations based on the set of structural selections.
Finally, we describe a set of general-purpose program transformations
that are provided in our current implementation.

\parahead{\little{}}

\noindent
To make the discussion of our design concrete, we choose to work with
a small functional language called \little{}, defined in \autoref{fig:syntax}.
A \little{} program is a sequence of top-level definitions, the last of which is
called \verb+main+. Notice that all (sub)expressions, (sub)patterns, definitions
(both at the top-level and locally via \verb+let+), and branches of \verb+case+
expressions are surrounded in the abstract syntax by orange boxes; these denote
\emph{code items} that will be exposed for selection and deselection in the user
interface. In addition, there are \emph{target positions}, denoted by blue
dots, before and after every definition, expression, and pattern in the program. Target
positions are ``abstract whitespace'' between items in the abstract syntax tree,
which will also be exposed for selection.

\begin{figure}[t]

\small

\centering


\vsepRule

\figSyntaxBegin

\figSyntaxRowLabel{Programs}{\varProgram}
  \targetBefore{\expDef{\varVar_0}{\varExp_0}{\targetsAround{\ \cdots\ }}}\
  \expTopDef{\mathtt{main}}{\varExp}
\figSyntaxSpaceNextCategory
\figSyntaxRowLabel{Expressions}{\varExpBare}
  \varConst
  \figSyntaxSpaceItem
  \varVar
  \figSyntaxSpaceItem
  \expFun{\varPat}{\varExp}
  \figSyntaxSpaceItem
  \expApp{\varExp_1}{\varExp_2}
  \figSyntaxSpaceItem
  \expCons{\varExp_1}{\varExp_2}
\figSyntaxLineBreak
\figSyntaxRow
  \expLet{\varPat}{\varExp_1}{\varExp_2}
  \figSyntaxSpaceItem
  \expCaseTwo{\varExp}{\expBranchWithId{\varExpId}{1}{\varPat_1}{\varExp_1}}{\cdots}
\figSyntaxSpaceNextCategory
\figSyntaxRowLabel{Patterns}{\varPatBare}
  \varConst
  \figSyntaxSpaceItem
  \varVar
  \figSyntaxSpaceItem
  \expNil
  \figSyntaxSpaceItem
  \expCons{\varPat_1}{\varPat_2}
\\[5pt]
\multicolumn{3}{c}{
  \mathrm{Expressions}\hspace{0.13in}\varExp\hspace{0.13in}{::=}\hspace{0.13in}\targetsAround{\wrapInBox{$\varExpBare$}}
  \hspace{0.30in}
  \mathrm{Patterns}\hspace{0.13in}\varPat\hspace{0.13in}{::=}\hspace{0.13in}\targetsAround{\wrapInBox{$\varPatBare$}}
}

\figSyntaxEnd


\caption{Syntax of \little{}.
The orange boxes and blue dots \mbox{identify} features for structural selection.}
\label{fig:syntax}
\end{figure}

\parahead{Code Tool Interface}

\begin{figure}[t]

\small


\begin{Verbatim}
  EditorState = { code: Program, selections: Set Selection }
  ActiveState = Active | NotYetActive | Inactive
  Options     = NoOptions | StringOption String
  Result      = { description: String, code: Program }

  CodeTool =
    { name : String
    , requirements : String
    , active : EditorState -> ActiveState
    , run : (EditorState, Options) -> List Result }
\end{Verbatim}
\caption{Code tool interface.}
\label{fig:code-tool-interface}
\end{figure}

\begin{figure*} 
\includegraphics[scale=0.40]{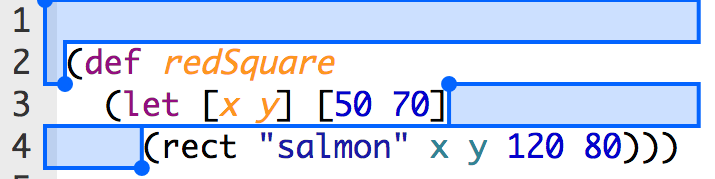}
\hspace{2em}
\includegraphics[scale=0.40]{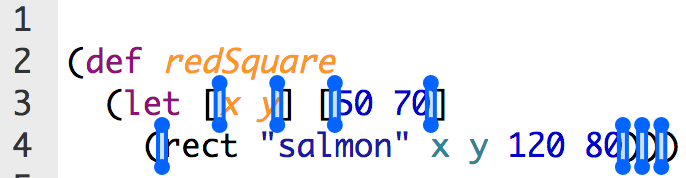}
\hspace{2em}
\includegraphics[scale=0.40]{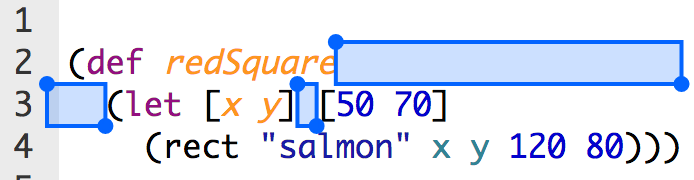}
\caption{Example target positions.}
\label{fig:whitespace-polygons}
\end{figure*}

Each code tool must implement the interface in \autoref{fig:code-tool-interface}.
A tool has access to the \verb+EditorState+, which contains a
\verb+Program+ and the \verb+Set+ of structural \verb+Selection+s within it.
Based on the \verb+EditorState+, the \verb+active+ predicate specifies
whether the tool is
\verb+Active+ (ready to run and produce \verb+Result+ options),
\verb+NotYetActive+ (could be \verb+Active+ if given more valid selections), or
\verb+Inactive+ (invalid based on the selections).
For example, \codeTool{Move Definitions} is \verb+NotYetActive+ if the user has
selected one or more definitions but no target position.
When invoked via \verb+run+, a tool has access to the \verb+EditorState+ and
configuration \verb+Options+, namely, an optional \verb+String+. This strategy
supports the ubiquitous \codeTool{Rename} tool. A more full-featured interface
may allow a more general set of configuration parameters; the challenge would be
to expose them using a lightweight user interface. In our implementation, all
transformations besides \codeTool{Rename} require \verb+NoOptions+. Each
\verb+Result+ is a new \verb+Program+ and a \verb+description+ of the changes.

This API between the user interface and code tool implementations is
shallow, in the sense that a code tool implementation can do whatever it
wants with the selection information. A framework for defining notions of
transformation correctness would be a useful line of work, but is beyond the scope of
this paper. Currently, code tools must be implemented inside the
\deuce{} implementation. Designing a domain-specific language for writing
transformations would be useful, but is also beyond the scope of this paper.

\parahead{Implementation in \sns{}}

We have chosen to implement our design within \sns{}~\citep{sns-pldi,sns-uist},
an interactive programming system for generating SVG images.
Whereas \sns{} provides capabilities for
directly manipulating the \emph{output} of a program, \deuce{}
provides capabilities for directly manipulating the
\emph{code} itself.

Direct code manipulation is particularly useful for a system like \sns{} for a couple reasons.
First, while the existing output-directed synthesis features in \sns{}
attempt to generate program updates that
are readable and which maintain stylistic choices in the existing code,
the generated code often requires subsequent edits, \eg{} to
choose more meaningful names, to rearrange definitions, and to
override choices made automatically by heuristics; \deuce{} aims to provide
an intuitive and efficient interface for performing such tasks.
Furthermore, by allowing users to interactively manipulate both code
and output, we provide another step towards the goal of \emph{direct
manipulation programming systems} identified by \citet{sns-pldi}.
These two capabilities---direct manipulation of code and
output---are complementary.

\sns{} is written in Elm (\url{http://elm-lang.org/}),
a language in which programs are
compiled to JavaScript and run in the browser.
The project uses the Ace text editor (\url{https://ace.c9.io/})
for manipulating \little{} programs.
(The second reason for the name \deuce{}
is that it extends Ace.)
We extended \sns{} to implement \deuce{};
our changes constitute approximately \locImplementation{} lines of
Elm and JavaScript code.
The new version (\version{0.6.2})
is available at \url{http://ravichugh.github.io/sketch-n-sketch/}.

\subsection{User Interface}
\label{sec:user-interface}

The goals of our user interface are, first, to expose \emph{structural
\maybecode{} selection widgets}---corresponding to the code items and target
positions in a \little{} program---and, second, to display an interactive menu
of active transformations based on the set of selections.

So that the additional features provided by \deuce{} do not intrude on
the text-editing workflow, we display structural selection widgets when hovering
over the code box only when the user is holding down the Shift key.
Hitting the Escape key at any time deselects all widgets
and clears any menus, returning the editor to text-editing mode.
This allows the user to quickly toggle between editing modes during
sustained periods in either mode.
When not using the Shift modifier key, the editor is a standard,
monospace code editor with familiar, unrestricted access to
general-purpose text-editing features.

\subsubsection{Structural \maybeCode{} Selection}

The primary innovation in our design is the ability to \emph{structurally
select} concrete source text corresponding to code items and target positions
from the abstract syntax tree of a program.

\parahead{Code Items}

Our
current 
implementation draws an invisible ``bounding polygon'' around the
source text of each expression, which tightly wraps the expression
even when stretched across multiple lines. These polygons serve as mouse
hover regions for selection, with the polygons of larger
expressions drawn behind the (smaller) polygons for the
subexpressions such that all polygons for child expressions partially
occlude those of their parents. Because complex expressions in
\little{} are fully parenthesized, it is always unambiguous
exactly where to start and end each polygon, and there are always
character positions that can be used to select an arbitrary
subexpression in the tree. Similarly, we create bounding polygons for
all patterns and definitions.

When hovering over an invisible selection polygon, 
\deuce{} colors the polygon to indicate that it has become the focus. Its
transparency and style is designed to resemble what might
otherwise be expected for text selection (\cf{} the screenshots in
\autoref{sec:overview}). Clicking a polygon selects the code item, making it visible even
after hovering away. Hovering the mouse back to the polygon and clicking it
again deselects the code item.

\parahead{Target Positions}

The user interface also draws polygons for the whitespace between code items for
selecting target positions.
\autoref{fig:whitespace-polygons}~(left) shows how our implementation draws
whitespace polygons slightly to the left of the beginning of a line, and until
the end of a line even if there are no characters on that line.
\autoref{fig:whitespace-polygons}~(center) shows whitespace polygons with
non-zero width even when there are no whitespace characters between adjacent code
items.

Another concern is that many target positions in the abstract syntax from
\autoref{fig:syntax} describe the \emph{same} space between code items. For
example, the expression
\expListTwo{\targetsAround{\mathtt{50}}}{\targetsAround{\mathtt{70}}}
on line 3 of
\autoref{fig:whitespace-polygons} contains both an after-\verb+50+ and before-\verb+70+
position. Because such target positions between adjacent items are redundant,
our implementation draws only one whitespace polygon. (This polygon is not
selected in any of the screenshots.)

A more interesting case is for the code items
\expTopDefNoBox
  {\targetsAround{{\varPat}}}
  {\targetsAround{{\varExp}}} and
\expLetNoBox
  {\targetsAround{{\varPat}}}
  {\targetsAround{{\varExp}}}
  {\cdots};
there is both an after-\varPat{} target and a before-\varExp{} target.
To allocate the whitespace between \varPat{} and \varExp{},
we take the following approach.
The space up to the first newline, if any, is dedicated to
after-\varPat{}; the remaining is for before-\varExp{}. If there is no
newline, then we do not expose any selection widget for before-\varExp{}.
For comparison, notice how the whitespace from the
end of line 2 to beginning of line 3 in
\autoref{fig:whitespace-polygons}~(right) is split into two polygons,
but the whitespace from the end of line 3 to the beginning of line 4 in
the \autoref{fig:whitespace-polygons}~(left) is not.
In other settings, it may be worthwhile to consider alternative approaches
to the design decisions above.

\subsubsection{Displaying Active Code Tools}

Several program transformations may be \verb+Active+ based on the items and
targets that are selected.
We design and implement a lightweight user interface for identifying, invoking,
and configuring \verb+Active+ transformations.

\parahead{Pop-up Panel}

When the user has entered structured editing mode (by pressing Shift) and selected at least one
item, we automatically display a menu near the selected items. The user
has already pressed a key to enter this mode, so our design does not require a
right-click to display the menu.
The user can drag the pop-up panel if it is obstructing relevant code.
We often manually re-positioned pop-up menus to make the screenshots
in \autoref{sec:overview} fit better in the paper.

\parahead{Hover Previews}

Each tool in the menu
has a list of \verb+Result+s, which appear
in a second-level menu when hovering the tool \verb+name+. Each second-level
menu item displays the \verb+description+ of the change, and hovering
over it previews the new \verb+code+ in the editor.
Clicking the item confirms the choice and clears all \deuce{} selections.
When there are few \verb+Result+s (\ie{} configuration options),
this preview menu provides a quick way to consider the options,
rather than going through a separate Configuration Wizard.
For tools that require multiple and non-trivial configurations,
however, the editor could fall back on separate, tool-specific
Configuration Wizards; our current implementation of \deuce{} does not support
this.

\subsection{Program Transformations}
\label{sec:little-transformations}

\begin{wrapfigure}{r}{0pt}
\includegraphics[scale=\lambdaScreenshotScale]{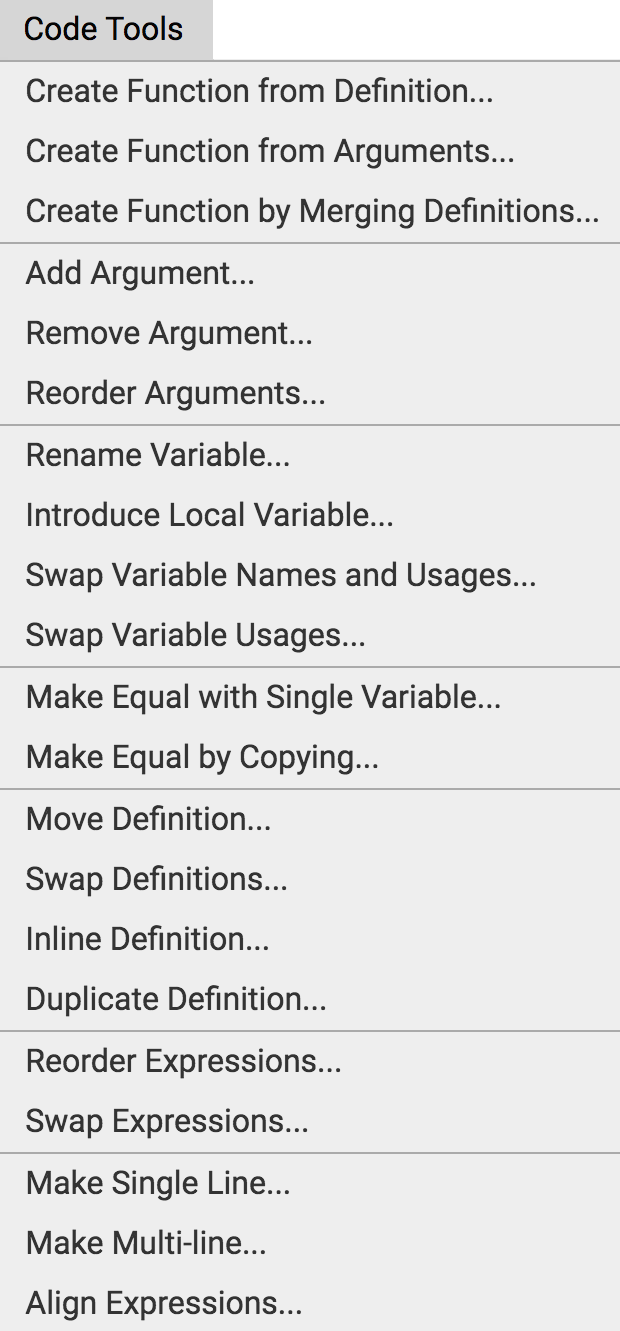}
\end{wrapfigure}
We have implemented a variety of program transformations, shown on the right.
While we believe these transformations form a
useful set of basic tools for common programming tasks, we
do not argue that these constitute a necessary or sufficient set.
One benefit of our design is that different sets of
transformations---such as
refactorings for class-based languages~\citep{Fowler1999},
refactorings for functional languages~\citep{Thompson2013},
transformations that selectively change program behavior~\citep{Reichenbach:2009}, and
task-specific transformations that do not have common, recognizable
names~\citep{Steimann:2012}---can be incorporated and
displayed to the user within our interface.

We limit our discussion below to transformations that
are not implemented in existing refactoring tools.
The \suppMaterials{} 
describe
other transformations, but these details are not necessary to understand the rest of the
paper.

\parahead{\codeTool{Make Equal with Single Variable}}

When multiple constants and an optional target position are selected,
the \codeTool{Make Equal with Single Variable} transformation introduces a new
variable, bound to one of the constants, and replaces all the
constants with the new variable. This has the effect of changing the
program to make each of these values equal.
The transformation attempts to suggest meaningful names, based on how the
selected expressions appear in the program.
For Example 1 from \autoref{sec:overview}, because the numbers \verb+120+ and
\verb+80+ are passed as the fourth and fifth arguments, respectively, to the
function \verb+(def rect (\(fill x y w h) ...))+, the suggested names include
\verb+w+ and \verb+h+.
The user is asked to choose a name.
The value itself (in this case, \verb+120+ or \verb+80+) is not as
important---the intention is that the values vary at once by a single
change---so, to keep the number of \verb+Result+s small,
the transformation does not ask the user to choose
which value to use for the variable.

\parahead{\codeTool{Move Definitions}}

Because of nested scopes and simultaneous bindings (\ie{} tuples), there are
many stylistic choices about variable definitions when programming in
functional languages.
The \codeTool{Move Definitions} transformation takes a set of selected definitions 
and a single target position, and attempts to move the
definitions to the target position. If the target position is before
an expression, a new \verb+let+-binding is added to surround the target.
Whitespace is added or removed to match the indentation of the target
scope. If the target position already defines a
list pattern, then the selected definitions are inserted into the list.
If the target position defines a single variable,
then a list pattern is created.
In cases where the intended transformation would capture variable uses or move
definitions above their dependencies (errors that are easy to make when using
text-edits alone), the transformation makes secondary edits (alpha-renaming
variables and moving dependencies) to the program to avoid these issues.
Our implementation of \codeTool{Move Definitions} also provides options for whether or not to
collapse multiple definitions into a single tuple, and also provides support for
rewriting arithmetic expression definitions as an alternative way to deal with dependency
inversion issues.

\section{User Study}
\label{sec:user-study}

We designed and implemented \deuce{} with the goal to incorporate structured
editing within a text-based program editor. In this section, we describe a user
study designed to measure the degree to which we were successful.

Besides the two novel mechanisms in our user interface design---structural
\maybecode{} selection and context-sensitive preview menus---that we wish to
evaluate, there are several additional factors at play. First, many users may
not have extensive experience with functional programming languages, especially
the custom \little{} language supported in our
implementation. Second, our implementation provides some familiar
transformations but some---particularly those involving target positions---are
not. Furthermore, some users may prefer to use text-editing rather than structured
edits, even when the latter can be used. These factors make it hard to perform a
direct comparison between our implementation of \deuce{} and an existing system,
such as Eclipse.

To mitigate these factors, we designed a study that compared \deuce{} with a
``baseline'' version of the system, with features designed to emulate the
traditional text-select-based interface described in \autoref{sec:intro}. We
then designed tasks, to be completed in both versions and \emph{without}
text-edits, to measure the effect of the new \deuce{} user interface features
compared to the baseline ones. Below, we describe the different configurations
of our system, our study procedures, and our results.

\subsection{System Configurations}

\newcommand{\startInteractionList}{}
\newcommand{\interactionItem}[2]{\paragraph{(#1) #2}}
\newcommand{\finishInteractionList}{}

\newcommand{\interactionName}[1]{Interaction {#1}}

Recall that tools may be \verb+Active+ or \verb+NotYetActive+ based on one or
more selected items and target positions (\autoref{fig:code-tool-interface}).

\paraheadNoDot{Traditional Mode (``Text-Select Mode'')}

To form the traditional mode of the tool, which we called Text-Select Mode in
the user study materials, we implemented four interactions separate from the
workflow described in \autoref{sec:overview} and \autoref{sec:user-interface} to
invoke code tools.

\startInteractionList{}

\interactionItem{A}{Code Tools Menu}

The editor displays a Code Tools menu at the top of the window with a list
of all transformations available in the system; this menu is akin to the Source
and Refactor menus in Eclipse. The user selects a tool from this menu without
first selecting anything in the program. Then, the editor displays a Tool
Configuration Panel that displays tool-specific instructions. Tool Configuration
Panels, which appear in all four interactions of Traditional Mode,
are discussed below.

\interactionItem{B}{Text-Select Single Argument + Code Tools Menu}

This interaction is like \interactionName{A}, except that the user first
text-selects an item or target in the code. Like Eclipse, text-selecting
requires the entire item to be selected, possibly with trailing or leading
whitespace. Our implementation provides more generous
text-selection mechanisms (\eg{} largest containing expression, smallest
surrounding expression), but the stricter version is used in the study because
it is more similar to existing approaches~\citep{Murphy-Hill-ICSE2008}. Also like
Eclipse, all tools are displayed and enabled in the Code Tools menu, even if the
tool is \verb+Inactive+ based on the selection.

\interactionItem{C}{Text-Select Single Argument + Right-Click Menu}

After first text-selecting an item, as in \interactionName{B}, the user right-clicks
to trigger a pop-up menu that displays \emph{only} plausible tools
(\verb+Active+ or \verb+NotYetActive+).
A similar workflow is provided by Eclipse.

\begin{wrapfigure}{r}{0pt}
\includegraphics[scale=\lambdaScreenshotScale]{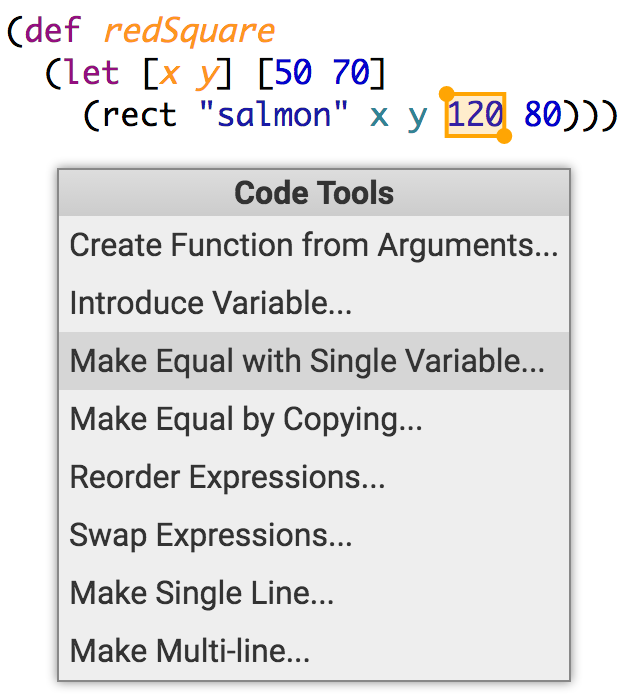}
\end{wrapfigure}
Returning to Example 1 from \autoref{sec:overview},
the screenshot on the right shows the right-click menu after
text-selecting the \verb+120+ constant.
By comparison, notice how this right-click menu displays more tools
(\verb+NotYetActive+ tools in addition to \verb+Active+ ones).
After the text selection is made, the editor draws an
orange box (as with \deuce{} widgets) to identify the selection.

\interactionItem{D}{Cursor-Select Single Argument + Right-Click Menu}

For atomic code items (\ie{} constants and variables), the user implicitly
selects the item by right-clicking on the token (rather than text-selecting it)
to trigger the right-click menu, as in \interactionName{C}. Again, a similar
workflow is provided by Eclipse.

\finishInteractionList{}

\paragraph{Tool Configuration Panels}

\begin{wrapfigure}{r}{0pt}
\includegraphics[scale=\lambdaScreenshotScale]{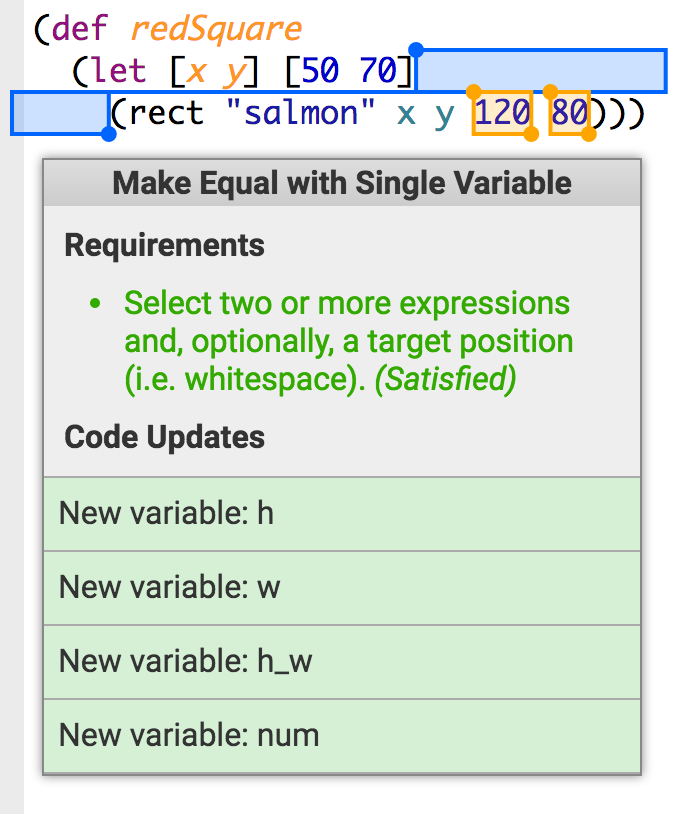}
\end{wrapfigure}
Each of the four interactions above trigger Tool Configuration Panels, which
display the \verb+requirements+ string that explains how to invoke the tool.
The user selects any additional arguments by hovering over and clicking
structural selection widgets. That is, structural selection widgets are
\emph{not} accessible to make the primary selection, but they \emph{are} used to
make all remaining selections in a Tool Configuration Panel.
The screenshot above shows the Configuration Panel after text-selecting
\verb+120+, as above, and then selecting \verb+80+ and a target position using
structural selection. Because the tool requirements are satisfied,
the panel displays the list of \verb+Result+s,
each of which can be hovered to preview the change before selecting it.

\paraheadNoDot{\deuce{} Mode (``Box-Select Mode'')}

This configuration, called Box-Select Mode in the user study materials, isolates
the new \deuce{} features.
To review, the user holds down the Shift key, then hovers over and clicks one or more
structural \maybecode{} selection widgets. When at least one widget is
selected, the pop-up preview menu displays the list of \verb+Active+ tools.

There is no Code Tools menu at the top of the editor in this mode, even
though the ``full'' version of our tool (not used by participants in the study)
does; the list of tool names and descriptions in Tool Configuration Panels (which
are \emph{not} accessible in \deuce{} Mode) can help understand unfamiliar
transformations.

\parahead{Combined Mode}

Our last configuration combines Traditional and \deuce{} Modes, with all
interactions described above.

\subsection{Questions and Procedures}

We sought to address several questions:

\begin{itemize}

\item Is either mode more effective for (a) completing tasks, (b) rapid editing, or (c) achieving more with fewer transforms?
\item Is either mode preferred by users? In which cases?

\end{itemize}

\noindent
To answer these questions, we designed the following IRB-approved, controlled
user study with \numUsers{} undergraduate and graduate students from the
University of Chicago.
We recruited users by sending emails to public mailing lists, offering
a monetary incentive of \$50 for participating in the two-hour study. Prior
experience with functional programming or \sns{} was not required. Each user
attended an individual session and was given the option to use the laptop and
mouse provided by us or their own devices.

The primary components of the study included a tutorial portion followed by a
tasks portion. We configured a pared-down version of the system that turned off
all \sns{} features unrelated to the interactions being studied. The tutorial
and tasks were set up as a self-guided progression of steps through the tool, to
be completed at the user's own pace.
In the description of the tutorial and tasks below, all random choices were
made independently of other choices, as well as across users.

Our system logged user events to analyze the tutorial and tasks. We also
recorded video of the users performing the tasks, for manual inspection
in situations where the log information was insufficient or
more difficult to process. Besides helping to get started and correct minor
issues unrelated to \deuce{}, the user study proctor did not answer any
questions about \deuce{} or the tasks.
To wrap up, users answered questions about their programming background and
experience using \deuce{} in an exit survey.

\begin{table*}
  \caption{Overview of the four head-to-head and two open-ended tasks.
  \#LOC is non-blank lines of code in the starting program.
  }
  \label{tab:tasks}
  \small
  \begin{tabular}{lccl}
    \toprule
    Name & \#LOC & \#Transforms &
    Example Tool Sequence (with minimum number of transforms required)\\
    \midrule
    One Rectangle & 9 & 3 & Swap Expressions; Move Definition; Swap Definitions\\
    Two Circles & 11 & 2 & Create Function from Definition; Reorder Arguments\\
    Three Rectangles & 11 & 2 & Creating Function by Merging Definitions; Rename\\
    Four Rings & 7 & 4 & Remove Argument; Rename; Move Definition; Add Arguments\\
    \midrule
    Four Squares & 9 & 7 & Create Function by Merging Definitions; Create Function from Arguments; Rename (5x)\\
    Lambda Icon & 10 & 8 & Make Equal with Single Variable (6x); Introduce Variable; Rename\\
    \bottomrule
  \end{tabular}
\end{table*}

\parahead{Tutorial}

The first part of the tutorial introduced ordinary text-based programming in
\little{}, emphasizing that the syntax would not be too important for subsequent
tasks.

The majority of the tutorial introduced the code tools using both
Traditional and \deuce{} Modes.
The first tool introduced---\codeTool{Rename Variable}, a familiar tool to
many---was explained using all five interaction modes. But because the four interactions
in Traditional Mode are largely similar, all subsequent tools introduced
in the tutorial had only one set of instructions for Traditional Mode. For all
tools introduced, a random choice was used to determine whether to explain
Traditional or \deuce{} Mode first. In total, 10 of the 22 code tools in
our implementation were demonstrated in the tutorial.
To give a flavor of the tutorial, Example 1 in \autoref{sec:overview}
is adapted from the steps that introduced the \codeTool{Make Equal} tools.
In addition to tool-specific tutorial steps,
we also dedicated a step for more practice with target positions, independent
of a specific tool, because the notion of target positions was likely to be
unfamiliar. 

\parahead{Tasks}

After the tutorial, users worked on six \emph{tasks}, each a
different program and a list of one or more edits to perform using code tools.
For some tasks, there were multiple different sequences of code tool invocations
that could lead to the desired result.
The starting programs ranged from 7 to 11 lines of code and required between
2 and 8 tool invocations (at minimum) to finish the tasks.
\autoref{tab:tasks} outlines the tasks. The Two Circles task was
presented as Example 2 in \autoref{sec:overview}. Extended task
descriptions
can be found in
the \suppMaterials{}.

Before every task, the participant was given a read-only reading period to understand the program
before seeing the list of edits to perform.
To emulate a real-world scenario where the programmer knows what to accomplish
but may not quite remember all the steps, the task directions were written in a
more natural style without direct reference to tool names---for example, ``move
the \verb+ring+ definition inside \verb+target+'' instead of ``invoke
\codeTool{Move Definition} on the \verb+ring+ definition with a target position
inside \verb+target+.''

Each of the first four tasks (``head-to-head tasks'') was performed twice, once
each in Traditional and \deuce{} Modes, resulting in eight \emph{trials}. The first four trials
comprised each of the four tasks, in random order and with one of the modes
randomly chosen per trial. For the next four trials, the order of tasks was, again,
randomized, each using the mode not chosen for the task in the first round.
After these eight trials, the user performed each of the last two tasks
(``open-ended tasks'') once using the Combined Mode---both Traditional and
\deuce{} Modes were available for use, to mix-and-match the two modes however
they saw fit.

For each task, comments showed what the desired final code should look like,
sometimes modulo minor whitespace differences. The editor provided an
indicator about whether the task was completed, giving the user the option to
Give Up at any point if needed. There was also a maximum time limit of six and
twelve minutes for each head-to-head and open-ended task, respectively, with no
indication about the time limit until and unless the user reached the two
and four minutes remaining mark, respectively.

\subsection{Results}
\label{sec:user-study-results}

Participants reported between 2 and 10 years of programming experience (mean: 5.1), of which
between 0 and 3 years involved functional programming (mean: 0.76). 10 participants (48\%)
reported no prior functional programming experience.
8 participants reported using tools that supported automated refactoring (Eclipse, IntelliJ, and
PyCharm all received multiple mentions). 4 participants reported some prior exposure to
previous versions of the \sns{} project, but none reported knowledge of the code tools presented
in the study.

For the study itself, 8 users brought their own laptop, the remaining 13 used
ours. 15 participants used a mouse, and 6 relied on their laptop's trackpad.
Each session took a mean of 1hr 44min (range: 1h 11m -- 2h 27m). Users spent between 23 and 66
minutes on the tutorial (mean: 41) and 20 and 65 minutes on the tasks (mean: 44). The remaining
time was spent on introductory remarks and the exit survey.
All users attempted all tasks. Two trials were discarded because of tool malfunction,
for a final total of 166 head-to-head trials and 42 open-ended tasks suitable for analysis.

\begin{figure}
\includegraphics[width=0.48\textwidth]{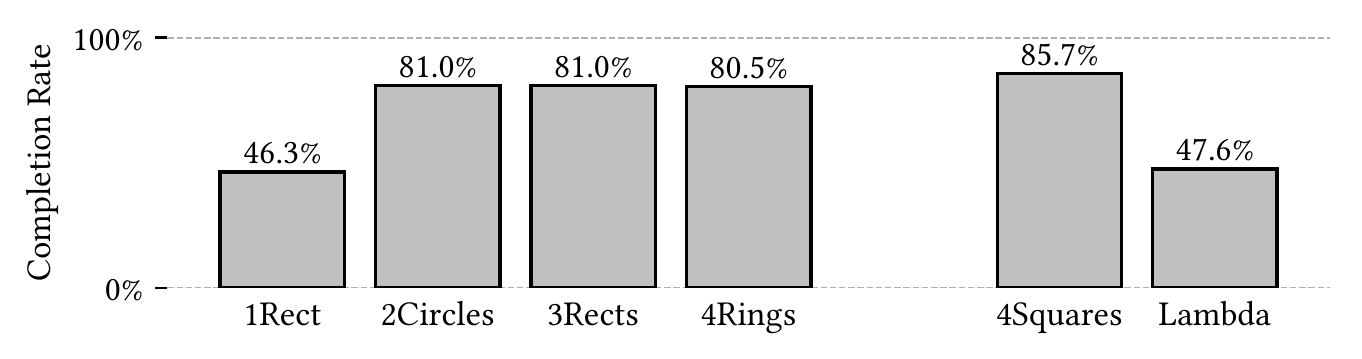}
\caption{Task completion rates pooled over both modes.}
\label{fig:pooled_completion_rates}
\end{figure}

The tasks proved moderately difficult. On average, each participant successfully completed 71\%
of the trials and open-ended tasks within the time limits, with 3 users
completing them all and 1 user failing to complete any.
\autoref{fig:pooled_completion_rates} shows completion rates by task.
The One Rectangle and Lambda tasks had particularly low completion rates. Based on
videos of failed attempts, many users struggled with choosing appropriate
tools---\eg{}~many chose \codeTool{Introduce Variables} rather than
\codeTool{Make Equal}, and some chose \codeTool{Inline} rather than
\codeTool{Move Definitions} in an attempt to create a tuple definition.
The tutorial was not sufficient for everyone to remember and understand all the
tools needed for the tasks.
The task descriptions may have also presented obstacles---\eg{}~for Lambda,
the phrase ``Define and use...'', along with \verb+(def [x y w h] ...)+ in the final
code, may have led some to use \codeTool{Introduce Variables}, which would then
require several roundabout transformations to complete the task.
We believe these difficulties
are largely independent of the user interface features.
We now address each of the research questions in turn.

\parahead{Is either mode more effective for completing tasks?}

\begin{figure}
\includegraphics[width=0.48\textwidth]{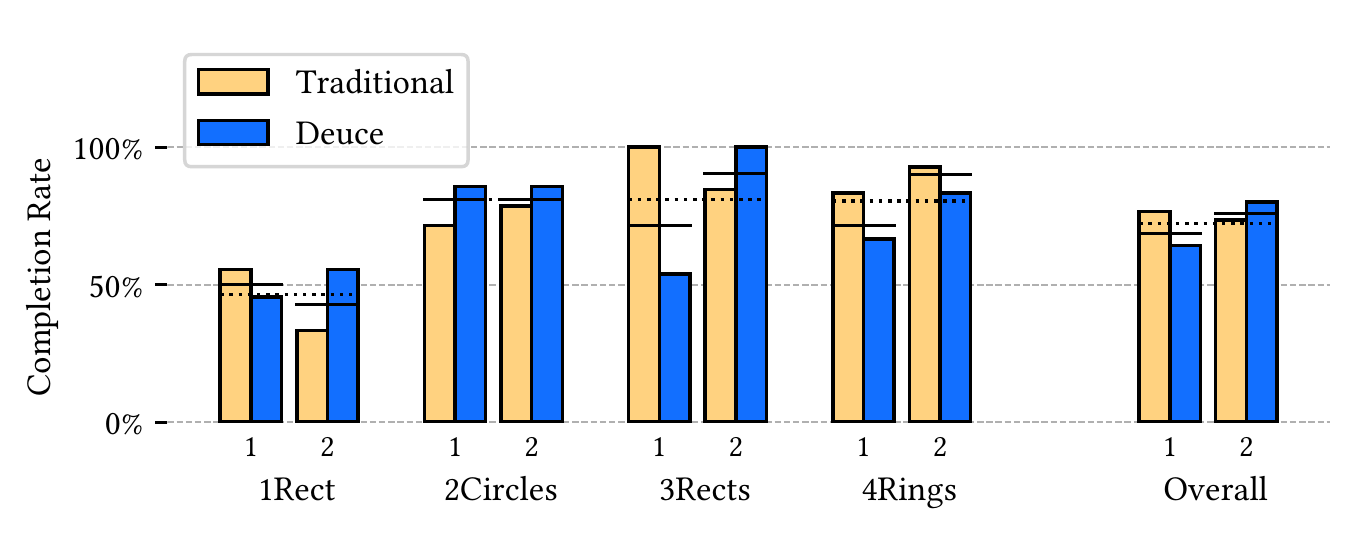}
\caption{Head-to-head task completion rates by mode and by subject's first/second encounter with task.
Overlaid lines indicated pooled completion rates.}
\label{fig:faceted_completion_rates}
\end{figure}

\autoref{fig:faceted_completion_rates} breaks down
completion rates for head-to-head tasks by mode. Because each was attempted
twice,
to assess possible learning effects from already completing a task in the other mode,
\autoref{fig:faceted_completion_rates} also differentiates
between the user's first or second encounter with each task.
Visually, the data suggest that on the first encounter with a task, Traditional Mode may
better facilitate completion, and is also a better teacher for the
subsequent encounter with
\deuce{} Mode. In contrast, a first encounter with \deuce{} Mode may be less helpful for the second
encounter
with Traditional Mode.

To control for learning effects, a mixed effects logistic
regression model~\cite{GelmanHill} was fit with \texttt{lme4}~\cite{lme4} to predict task completion probability based upon fixed effect
predictors for the mode (coded as 0 or 1), the trial number (1-8), whether the trial was the second
encounter with the task (0 or 1), whether the participant used a mouse (0 or 1), whether the
participant used their own computer (0 or 1), and the interaction of mode with the second
encounter (0, or 1 when \deuce{} Mode and a second encounter). To model
differences in user skill and task difficulty, a random effect was added for each participant as well as
each task, and a random interaction was added to model
differences in the second encounter difficulty per task. Reported p-values are based on Wald Z-statistics.

In the fit model, the coefficient for mode was on the edge of significance (p=0.057), indicating
that Traditional Mode did better facilitate task completion on the first encounter with a task.
Given this, \deuce{} Mode performed better than expected on the second encounter (interaction term
p=0.036), but not enough to confidently say that \deuce{} Mode was absolutely better than
Traditional Mode for the second encounter (p=0.17). No other fixed effect coefficients approached significance.

\deuce{} Mode therefore seems to present a learning curve, but may be just as effective as
Traditional Mode once that learning curve is overcome. This interpretation accords with the surveys:
5 participants wrote that Traditional Mode might be better for learning, and
4 participants---including 3 of the previous 5---said \deuce{} Mode was
better when they knew the desired transformation.
However, the data may be
alternatively explained if \deuce{} Mode on the first encounter is a poor teacher, actively
misleading users on the second encounter with Traditional Mode.

\begin{figure}
\includegraphics[width=0.48\textwidth]{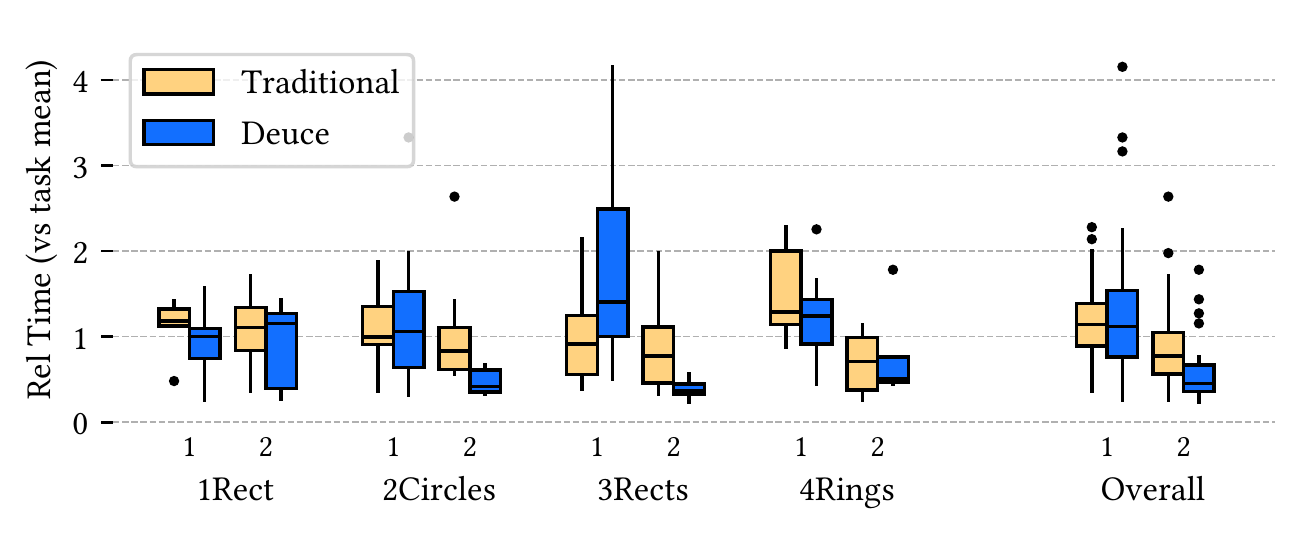}
\caption{Head-to-head task durations for successfully completed trials, scaled relative to the mean
time per task.}
\label{fig:task_times}
\end{figure}

\paraheadNoDot{Is either mode more effective for rapid editing?} Among trials successfully
completed, the duration of each trial was measured from the start of configuration of the first refactoring
to the end of the final refactoring. The distribution of these timings is presented
in \autoref{fig:task_times}, scaled relative to the mean duration for each task.

Again, to tease out if any of these differences are significant, from the same predictors described above
two linear mixed effects models were fit to predict (1) trial duration and (2) the logarithm of trial duration
(\ie{} considering effects to be multiplicative rather than additive).
Percentile bootstrap p-values
for the fixed effect coefficients were calculated
from 10,000 parametric simulate-refit samples.\footnote{See \url{https://www.rdocumentation.org/packages/lme4/versions/1.1-13/topics/bootMer}}
For the first encounter
with a task, Traditional Mode was insignificantly faster (by 13 seconds, p=0.44; or 9.2\%, p=0.52).
However, \deuce{} Mode was
on average 25 seconds (p=0.13) or 36\% (p$<$0.01) faster for the second encounter with a task, suggesting that
\deuce{} Mode may be faster once users become familiar with the available tools. Most of the gain
comes from less time spent in configuration---after discounting all idle thinking time between
configurations, the model still reveals an 18 second difference.

\paraheadNoDot{Is either mode more effective for achieving more with fewer transforms?} To
determine if either mode facilitated more efficient use of interactions, the
same mixed effects model was fit to predict the number of refactorings invoked during each
successful trial, as well as the number of Undos.
On the first encounter with a task, Traditional Mode accounted for an average
of 2.0 fewer refactorings (p$<$0.01) and 2.1 fewer Undos (p$<$0.01), but on the second encounter no significant difference in
number of refactorings or Undos was indicated. As a second encounter with \deuce{} Mode is faster than
Traditional Mode, the speed gain thus appears to be explained by faster invocations rather than fewer
invocations.

\paraheadNoDot{Is either mode preferred by users? In which cases?}

\begin{figure}
\includegraphics[width=0.48\textwidth]{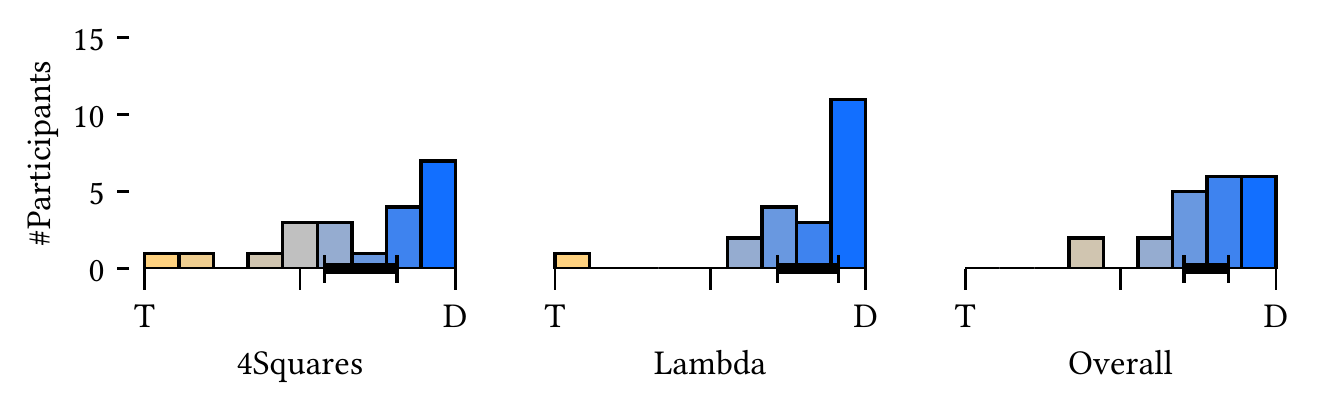}
\caption{Distribution of user preferences for Traditional vs. \deuce{} Modes as measured by the
ratio of refactorings performed by the user in each mode on the open-ended tasks. Far left
represents all Traditional Mode refactorings; far-right indicates all \deuce{} Mode refactorings.
The 95\% confidence interval for the mean preference across all users is indicated (via percentile bootstrapping, 10,000 samples).}
\label{fig:objective}
\end{figure}

\begin{figure}
\includegraphics[width=0.48\textwidth]{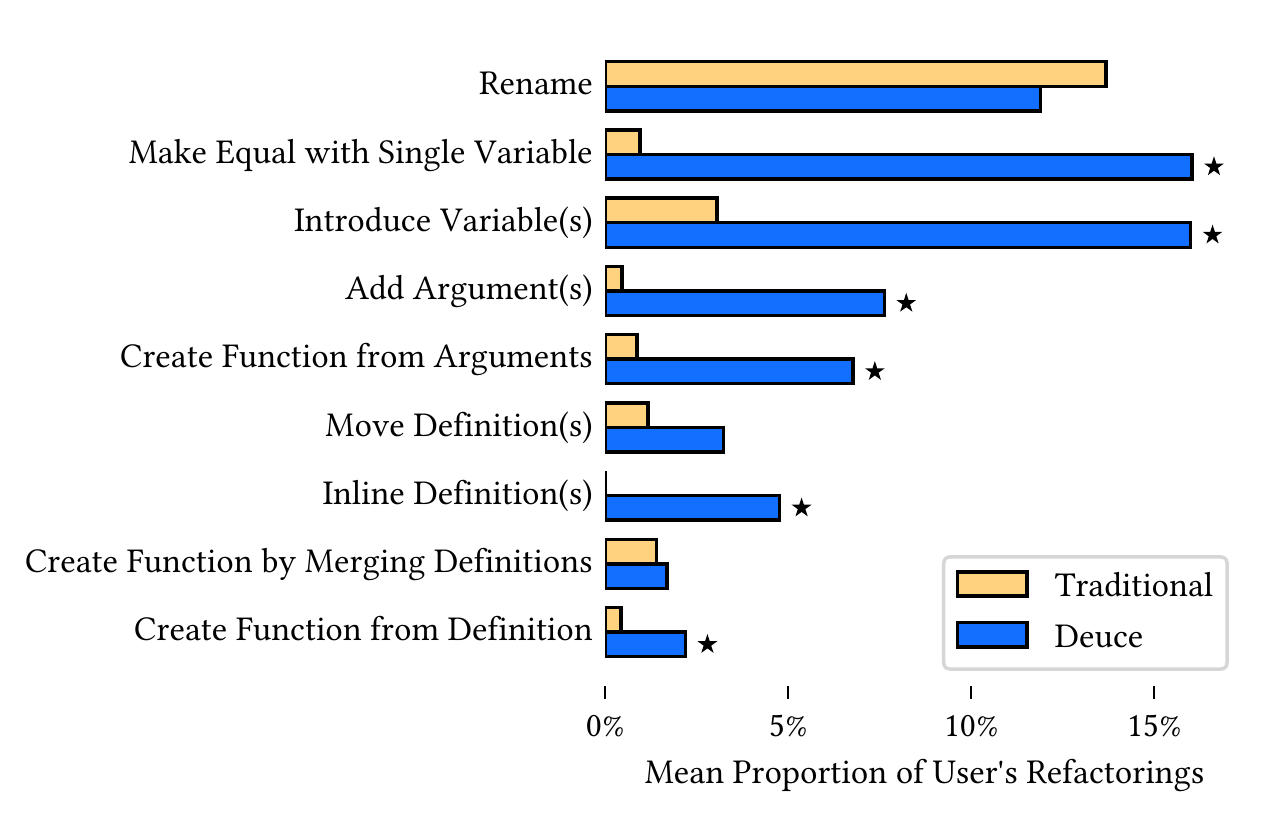}
\caption{Mode usage for tools used by at least half of participants on the open-ended tasks. Deuce mode is preferred for most tools.
Stars indicate differences significant at the 95\% level (via percentile bootstrapping, 10,000 samples).}
\label{fig:tool_modes}
\end{figure}

The two final open-ended tasks allowed participants to mix-and-match the two modes
as they pleased. As shown in \autoref{fig:objective}, on both tasks the overwhelming number of users performed a
greater share of refactorings using \deuce{} Mode. We believe a main advantage of \deuce{}
 Mode is that it simplifies the configuration of refactorings that require multiple arguments, as the user
may select all the arguments together before choosing a transformation from a short menu. In
Traditional Mode, the workflow is stuttered: the user must select a single argument, right-click
to choose a transformation, then select the remaining arguments. However, for a refactoring requiring
only a single argument, Traditional Mode is more streamlined: a user may simply select the desired
transformation immediately after right-clicking on the first argument. Thus, for single-argument
refactorings, \deuce{} Mode's advantages may be limited. A breakdown of mode usage by popular tools
(\autoref{fig:tool_modes}) lends support to this hypothesis. For the most commonly used tool, \codeTool{Rename},
which always takes only a single argument, participants used Traditional and \deuce{} Modes with
roughly even frequency. Most other tools showed strong preferences towards \deuce{} Mode, with
the notable exception of  \codeTool{Create Function by Merging Definitions}.
Because the Four Squares task required invoking this tool with four expressions, according to
the hypothesis, users should prefer \deuce{} Mode. The videos revealed that several
users were unable to discover how to structurally select a function call, which required hovering
on the open parenthesis (not demonstrated
in the tutorial). Several of these users were, however, able to invoke the tool by
text-selecting a function call or by starting from the full Code Tools menu.

\begin{figure}
\includegraphics[width=0.48\textwidth]{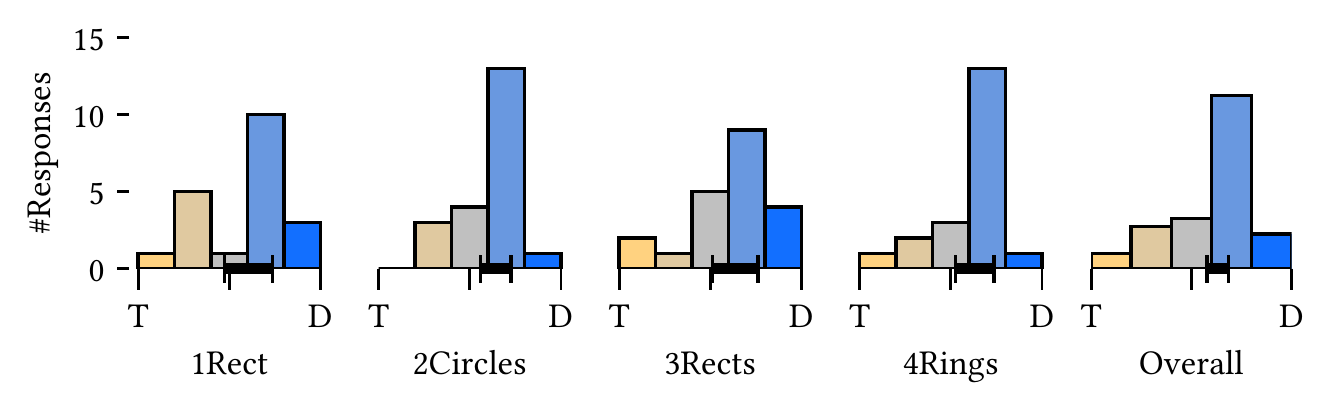}
\caption{Surveyed subjective preference for Traditional vs. \deuce{} Modes for the head-to-head tasks.
The 95\% confidence interval for the mean preference across all users is indicated (via percentile bootstrapping, 10,000 samples).}
\label{fig:subjective}
\end{figure}

Subjectively, the concluding survey asked whether \deuce{} or Traditional Mode worked better for
each head-to-head task, measured on a 5-point scale from ``Text-Select Mode worked much better'' to
``Box-Select Mode worked much better''. For each participant, a random choice
determined which mode appeared at each end of the scale.
As shown in \autoref{fig:subjective}, on average a similar
modest preference for \deuce{} Mode was expressed for each task.

On the free-response portion of the survey, several explanations were given for this
preference for \deuce{} Mode.
3 participants
appreciated the ability to select multiple arguments;
2 other participants
appreciated selecting all arguments before selecting a tool;
1 other participant
appreciated the smaller menu of refactorings; and
1 other participant
appreciated the ease of starting a refactoring by clicking code objects rather
than having to create a text selection.

Altogether, users demonstrated a strong objective and modest subjective preference for
\deuce{} over Traditional Mode, suggesting that \deuce{} accomplishes its goal
to provide a more human-friendly interface to identify, configure, and invoke refactorings.

\parahead{Limitations}

There are several threats to the validity of our experimental setup.
One is that our emulation of traditional features may have been less
effective than those features in existing tools. Another is that
the participants may have felt compelled to use \deuce{} Mode (which could
likely have been deduced to be more novel than Traditional Mode) more during the
open-ended tasks---and pronounce a preference for it in the survey---because the
participants were drawn from the same academic community as the authors. Another
is that participants used the tool in heterogeneous environments---different
computers and browsers, configured with different screen sizes and mouse
settings. Performance on the tasks may have also been affected by the presence
of the user study proctor and video recording device. According
to self-reported assessments, participants were relatively unfamiliar with
functional programming and with refactoring tools, so the results may differ
for users with more extensive experience. Finally, our results were obtained on
small programs and tasks in a prototype language.

\parahead{Future Improvements}

There are opportunities to improve our implementation of \deuce{}.
First, to reduce the learning curve, it would be worth adding more explanatory
features (\eg{}~in a tutorial, or within the tool when the user selects certain
kinds of items for the first time), particularly for unfamiliar transformations
(\eg{}~\codeTool{Move Definitions}) and for unfamiliar user interface features
(\ie{}~target positions). Enabling the full Code Tools menu may also help
because of the descriptions of requirements in the Tool Configuration Panels
(\cf{}~the ``\deuce{} Mode'' discussion). Also, to allow easy corrections of
misconfigured refactorings, it would help if
Undo restored the previous selection state rather than just the previous version of
the code; we have since implemented this feature.

\section{Related Work}
\label{sec:discussion-related}

We describe the most closely related ideas in structured
editing and refactoring.
\citet{Barista}, \citet{DNDRefactoring}, and \citet{HazelnutSNAPL}
provide more thorough introductions.

\subsection{User Interfaces for Structured Editing}


Compared to traditional text-selection and menus, several alternative user
interface features have been proposed to integrate structured editing more
seamlessly within the text-editing workflow.

\parahead{Text Selection}


\citet{Murphy-Hill-ICSE2008} identify that text selection-based refactoring is
prone to error, particularly for statements that span multiple lines and that
have irregular formatting.
They propose two prototype user interface mechanisms, called
Selection Assist and Box View, to help.
With Selection Assist, the user positions the cursor at the start of a statement, and
the entire statement is highlighted green to show what must be
selected (using normal text-selection).
With Box View, the editor draws a separate panel (next to the code editor) that
shows the tree structure of the program with nested boxes. When selecting text
in the editor, the nested boxes are colored according to which code items are
completely selected. Similarly, the user can select a nested box in the Box View
to select the corresponding text in the code.

In contrast, our structural selection polygons are drawn
directly atop the code, at once helping to identify (like Box View) and select (like
Selection Assist), which aims to mitigate the context switching
overhead of Box View identified by \citet{Murphy-Hill-ICSE2008}.



\parahead{Drag-and-Drop Refactoring}
\label{sec:discussion-related-dndr}

\citet{DNDRefactoring} propose a tool called \dndr{} that eliminates the use
of menus altogether. They demonstrate how many common Eclipse
refactorings can be unambiguously invoked with a drag-and-drop gesture
without the need for any additional configuration.
This is a compelling workflow for situations
in which the user can (a) readily \emph{identify} an intended
refactoring based on a preconceived notion (\eg{} its name), (b)
unambiguously \emph{invoke} the intended refactoring by a single-source,
single-target drag-and-drop gesture, and (c) accept the \emph{default configuration}
of the refactoring.
It would be useful to add drag-and-drop gestures to \deuce{}
for transformations that satisfy these three conditions.
%
However, our user interface supports situations when one or more
of these three conditions fails to hold.

\parahead{Hybrid Editors}

Compared to ``fully'' structured editors, several \emph{hybrid editor}
approaches augment text-based programs with additional information.
Barista~\cite{Barista} is a hybrid Java editor where
\emph{structure views} can be implemented to present
alternate representations of structural items instead of text. For
example, an arithmetic expression may be rendered with mathematical
symbols, a method may be accompanied by
interactive documentation with input-output examples, and structures
may be selectively collapsed, expanded, or zoomed.
\citet{ActiveCodeCompletion} introduce a similar notion to
structure views, called \emph{palettes}, where custom displays can be
incorporated based on the type of a subexpression. For example, a
color palette can provide visual previews of different candidate color
values, and a regular expression palette can show input-output examples
for different candidate regular expressions.
In Greenfoot~\cite{Greenfoot}, program text is separated into
structural regions called \emph{frames}, which are created and
manipulated with text- and mouse-based operations that are orthogonal
to the text-edits within a frame.
Code Bubbles~\cite{CodeBubbles1} allows text fragments to
be organized into \emph{working sets}, which are collections of code,
documentation, and notes from multiple files that can be organized in
a flexible way.
Outside of the views, palettes, frames, and working sets in the
above hybrid editors, the user has access to normal text-editing tools.

Our approach is complementary to all of the above: in places where
code fragments---regardless of their granularity and their
relationship to alternative or additional pieces of information---are
represented in plain text, we aim for a lightweight user interface to
structurally manipulate it.

\parahead{Refactoring with Synthesis}

In contrast to direct manipulation in \dndr{} and \deuce{},
\citet{RefactoringWithSynthesis} propose a
workflow where the user starts a refactoring with
text-edits---providing some of the changes after the refactoring---and
then asks the tool to synthesize a
sequence of refactorings that complete the task.
This text-based interface and the mouse-based interfaces of \dndr{}
and \deuce{} are complementary.


\subsection{Program Transformations}


Automated support for
refactoring~\cite{GriswoldThesis,Fowler1999,SmalltalkRefactoring} has been aimed
primarily at programs written in class-based, object-oriented languages.


\parahead{Refactoring for Functional Languages}

\hare{}~\cite{Thompson2013,HaReThesis1,HaReThesis2}
is a refactoring tool for functional languages, such as Haskell,
where features---including first-class functions
(\ie{} lambdas), local bindings, tuples, algebraic datatypes, and type
polymorphism---lead to editing tasks that are different from those
supported in most typical refactoring tools for object-oriented programs.
Our user interface could be incorporated by \hare{} to expose the
supported transformations with lightweight direct manipulation.
\hare{} provides a larger catalog of transformations than our
current implementation of \deuce{}. However, the details of our \codeTool{Move
Definitions} and \codeTool{Make Equal} transformations are, to the best of our
knowledge, not found in existing tools.

\section{Conclusion}


Based on our experience and the results of our user study,
we believe \deuce{} represents a proof-of-concept for how to achieve a
lightweight, integrated combination of text- and structured editing. In future
work, our design may be adapted and implemented for full-featured programming
languages and development environments, incorporating additional well-known
transformations (\eg{}~\citet{Fowler1999,Thompson2013}). Additional direct code
manipulation gestures, as well as incremental parsing (\eg{}~the algorithm
of \citet{Wagner:1998} used by Barista~\citep{Barista}),
could further help streamline, and augment,
support for structured editing within an unrestricted text-editing workflow.

\begin{acks}
The authors thank Shan Lu, Elena Glassman, Aaron Elmore, Peter Scherpelz,
and Blase Ur for suggestions about this paper.
This work was supported by
\grantsponsor{GS100000001}{National Science Foundation}{http://dx.doi.org/10.13039/100000001}
Grant No.~\grantnum{GS100000001}{1651794}, and
a University of Chicago Liew Family Research Fellows Grant.
\end{acks}


\clearpage

\onecolumn

\appendix

\section{Program Transformations}
\label{sec:appendix-transformations}

In this section, we supplement the discussion in
\autoref{sec:little-transformations} of transformations currently implemented in
\deuce{}.

\subsection{General-Purpose Transformations}

\begin{figure*}[b]
\small

\newcommand{\moreName}[1]{\textit{#1}}

\beginToolTable
\toolTableHeaders
\toolCategoryReuse
\rowTool
  {Create Function from Definition}
  {0}{1}{1}{0}
  {all constants or only named constants}
\rowTool
  {Create Function from Arguments}
  {1+}{0}{0}{0}
  {}
\rowTool
  {Merge}
  {2+}{0}{0}{0}
  {}
\hline
\toolCategoryStyle
\rowTool
  {Move Definitions}
  {0}{1+}{1+}{1}
  {renaming; dep. lifting; dep. inversion}
\rowTool
  {Swap Definitions}
  {0}{2}{2}{0}
  {}
\rowTool
  {Introduce Variable}
  {1+}{0}{0}{1}
  {}
\rowTool
  {Add Arguments}
  {1+}{0}{0}{1 in arg list}
  {}
\rowTool
  {Remove Arguments \moreName{at function definition}}
  {0}{1+ arg}{0}{0}
  {}
\rowTool
  {Remove Arguments \moreName{at function call}}
  {1+ arg}{0}{0}{0}
  {}
\rowTool
  {Reorder Arguments \moreName{at function definition}}
  {0}{1+}{0}{1 in arg list}
  {}
\rowTool
  {Reorder Arguments \moreName{at function call}}
  {1+}{0}{0}{1 in arg list}
  {}
\rowTool
  {Reorder List Items}
  {1+}{0}{0}{1 in parent list}
  {}
\rowTool
  {Rename Variable \moreName{at definition}}
  {0}{1}{0}{0}
  {new name}
\rowTool
  {Rename Variable \moreName{at use}}
  {1 variable}{0}{0}{0}
  {new name}
\rowTool
  {Swap Variable Names and Usages}
  {0}{2}{0}{0}
  {}
\rowTool
  {Inline Definition}
  {0}{1+}{0}{0}
  {}
\rowTool
  {Duplicate Definition}
  {0}{1+}{0}{1}
  {}
\rowTool
  {Clean Up}
  {0}{0}{0}{0}
  {}
\rowTool
  {Make Single Line}
  {1}{0}{0}{0}
  {}
\rowTool
  {Make Multi-Line}
  {1}{0}{0}{0}
  {}
\rowTool
  {Align Expressions}
  {2+}{0}{0}{0}
  {}
\hline
\toolCategoryChange
\rowTool
  {Make Equal with Single Variable}
  {2+ constants}{0}{0}{0}
  {choose suggested name}
\rowTool
  {Make Equal by Copying}
  {2+}{0}{0}{0}
  {choose expression to copy}
\rowTool
  {Reorder Expressions}
  {1+}{0}{0}{1}
  {}
\rowToolNoBreak
  {Swap Variable Usages}
  {0}{2}{0}{0}
  {}
\end{tabular}

\caption{General-purpose transformations in \deuce{}.}
\label{fig:transformations-1}
\end{figure*}

\autoref{fig:transformations-1} shows a list of the general-purpose code tools
in our implementation, along with the number of selected code items and target
positions required to make the tools \verb+Active+.

\parahead{\codeTool{Move Definitions}}

A common, mundane text-editing task is to rearrange definitions. While
conceptually simple, there are several aspects that are subject to
error, such as making sure not to break dependencies and making sure not to break
scoping. Furthermore, particularly in functional languages where
local-bindings can be arbitrarily nested and where tuples can be used to
simultaneously define multiple bindings, there are many stylistic
reasons for arranging definitions in a certain way. These choices are
often in flux during the prototyping and repairing process, where
definitions may be reordered to more clearly explain program
dependencies and to ensure that the concrete layout fits nicely within
the screen width.

The \codeTool{Move Definitions} transformation takes a set of selected patterns
and a single target position, and attempts to move the pattern and its
definition to the target position. If the target position is
an expression, a new let-binding is added to surround the target.
Whitespace is added or removed to match the indentation of the target
scope. If the target position already defines a
list pattern, then the selected definitions are inserted into the list.
If the target position defines a single variable,
then a list pattern is created.

\begin{figure*}[t]

\newcommand{\scaleABD}{0.29}
\newcommand{\scaleC}{0.41}
\newcommand{\vsepFigure}{\vspace{0.08in}}

\begin{subfigure}{\textwidth}
  \centering
  \includegraphics[scale=\scaleABD]{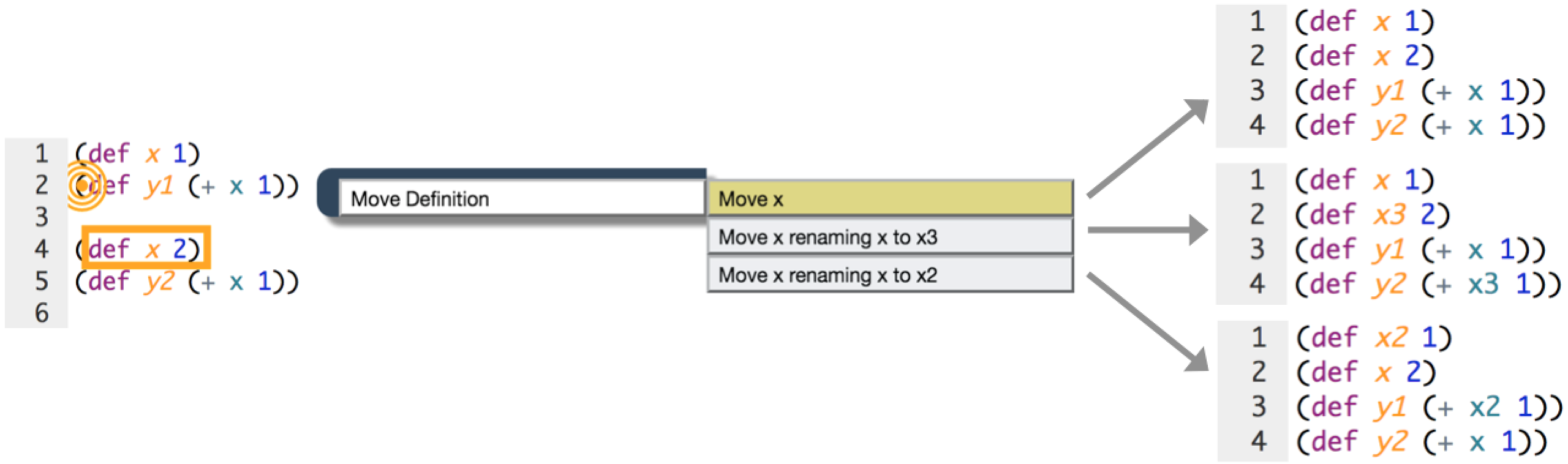}
  \caption{Simply moving the definition of \texttt{x} on line 4 to line 2 would
change the binding structure of the program. In addition to this
option, so the second and third options use renaming to
preserve the binding structure of the original program.}
  \label{fig:move-def-a}
\end{subfigure}

\vsepFigure

\begin{subfigure}{\textwidth}
  \centering
  \includegraphics[scale=\scaleABD]{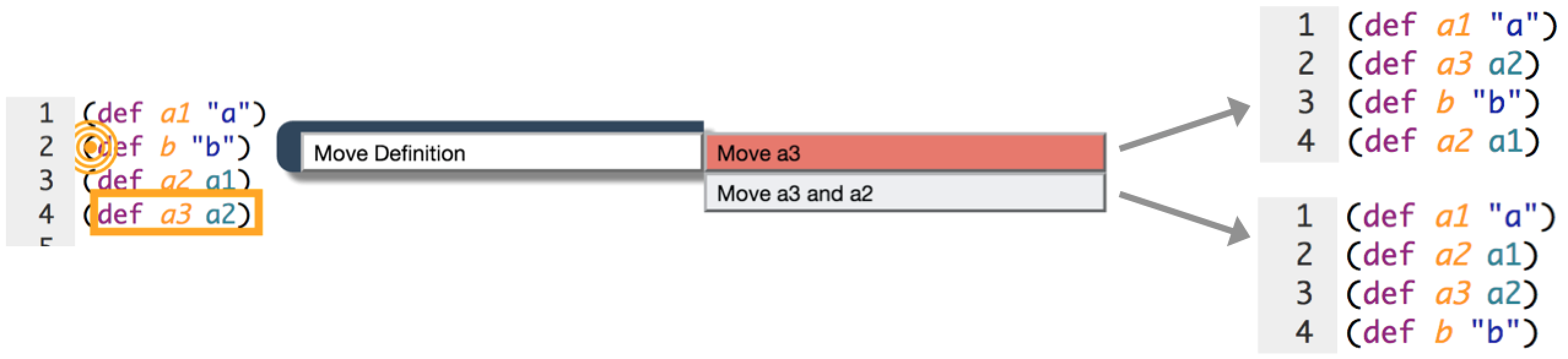}
  \caption{Simply moving the definition of \texttt{a3} above \texttt{b} would
result in a use of \texttt{a2} before it is defined, so the second
option additionally lifts the definition of \texttt{a2}.}
  \label{fig:move-def-b}
\end{subfigure}

\vsepFigure

\begin{subfigure}{\textwidth}
  \centering
  \includegraphics[scale=\scaleC]{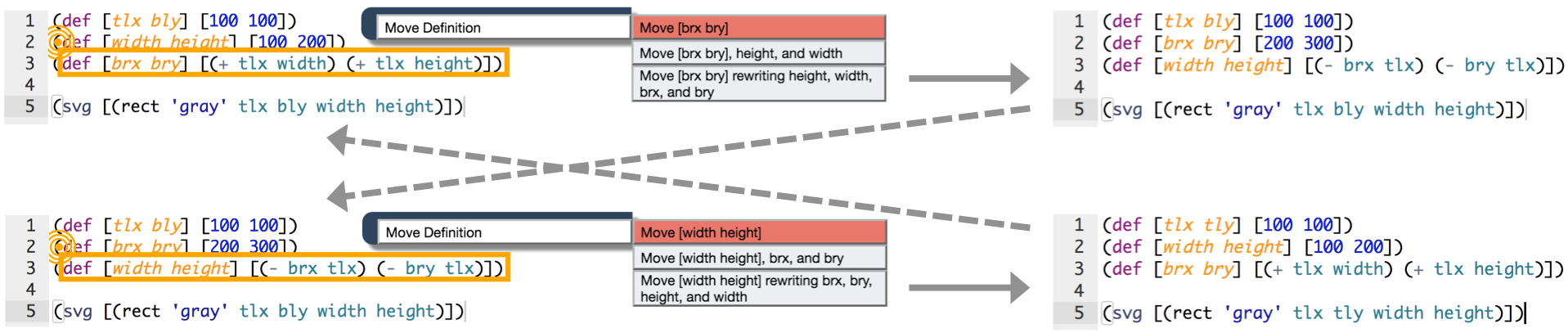}
  \caption{For arithmetic constraints among expressions---such as the
top-left corner, width, height, and bot-right corner of a rectangle---
\codeTool{Move Definitions} can move definitions above dependencies by rewriting
the expressions.}
  \label{fig:move-def-c}
\end{subfigure}

\vsepFigure

\begin{subfigure}{\textwidth}
  \centering
  \includegraphics[scale=\scaleABD]{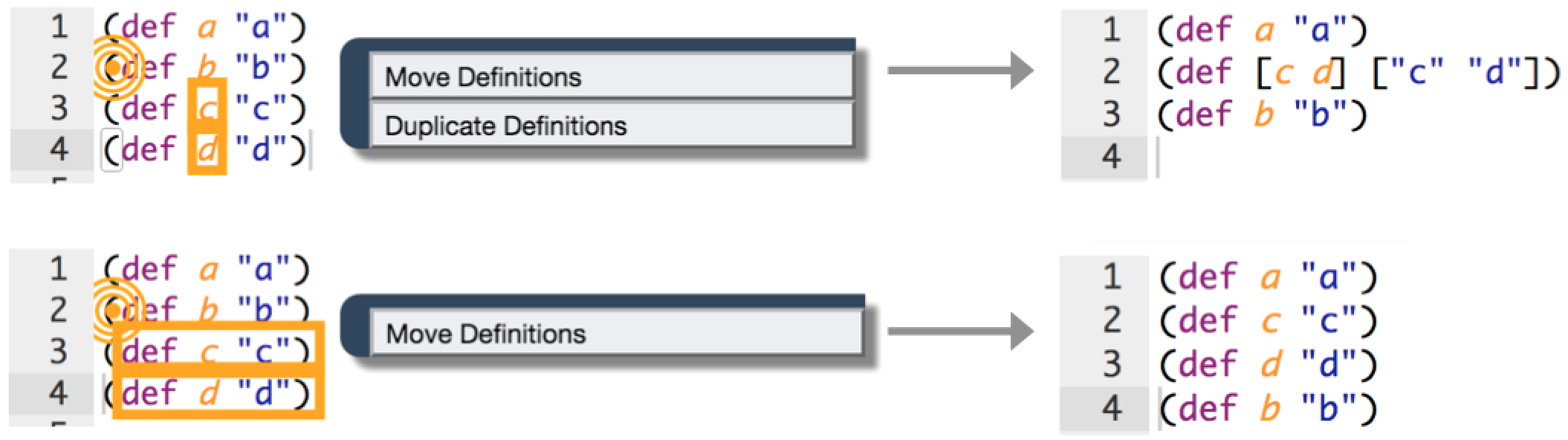}
  \caption{When selecting the patterns \texttt{c} and \texttt{d} and
moving them before \texttt{b}, they are put into a single definition.
When selecting the definitions \texttt{c} and \texttt{d}, the separate
definitions are preserved in the new location.}
  \label{fig:move-def-d}
\end{subfigure}

\caption{Examples demonstrating four configuration options for
\codeTool{Move Definition} transformation.}
\label{fig:move-def}
\end{figure*}

There are three cases in which \deuce{} provides the user options for
how to correct an otherwise invalid transformation.
In all cases, the user may choose the original
invalid option since breaking the code temporarily may be the
intention during the course of prototyping and repairing.
Note that the screenshots in \autoref{fig:move-def} show a preliminary
implementation of \deuce{}, with cosmetic differences in the user interface.

\paragraph{Option: Renaming to Preserve Binding Structure}

One issue is that the binding structure may change. For example, in
\autoref{fig:move-def-a}, the uses of \verb+x+ in the expression
\verb&(+ x 1)& on lines 2 and 5 resolve to different definitions of
\verb+x+, on lines 1 and 4, respectively. Moving the definition of
\verb+x+ from line 4 before \verb+y1+ will result in a program that
evaluates safely but with different binding structure.  In this case,
the \codeTool{Move Definition} transformation provides several options:
the unsafe option that performs the transformation without renaming
which allows a binding to be captured, and two safe options that
rename either of the definitions to avoid capture.

\paragraph{Option: Lifting Dependencies}

A second potential issue is that the definitions would be moved before
its dependencies. In the example in \autoref{fig:move-def-b},
\verb+a2+ is defined to be \verb+a1+, and \verb+a3+ is defined to be
\verb+a2+. Trying to move \verb+a3+ above \verb+b+ is not safe,
because \verb+a2+ is not in scope. The transformation provides two
options: the unsafe transformation, and
a version where the dependency, \verb+a2+, is automatically
moved as well to make the original transformation safe.

\paragraph{Option: Inverting Dependencies}

A third situation is when the user may want to rewrite definitions so
that a dependency violation can be avoided. In the top-half of
\autoref{fig:move-def-c}, the program defines the top-left point
(abbreviated to \verb+tl+ for space) to be \verb+(100,100)+ and the
\verb+width+ and \verb+height+ to be \verb+100+ and \verb+200+,
respectively; the location of the bottom-right point (abbreviated to
\verb+br+) is derived in terms of these parameters. Dragging the
third definition above the second results in \verb+brx+ and \verb+bry+
being rewritten to constants that the previous expressions evaluated
to (\ie{} \verb+200+ and \verb+300+),
and then \verb+width+ and \verb+height+ are rewritten to arithmetic
expressions that preserve the original relationship
(\ie{} \verb&(- brx tlx)& and \verb&(- bry tly)&). In situations
where there are constraints among expressions---particularly common in
programming domains that generate visual output, such as
a web application or data visualization---this transformation
allows the programmer to vary the choices about which parameters are
defined first in the program with constants and which are derived
in terms of them. The bottom half
of \autoref{fig:move-def-c} shows how the new definitions
can also be inverted, returning the program to the original.

\paragraph{Option: Flattening or Preserving Definition Structure}

As listed in \autoref{fig:transformations-1}, \codeTool{Move Definition} can be
initiated by selecting either one or more patterns (\eg{} just the
variables \verb+c+ and \verb+d+ in the top half of \autoref{fig:move-def-d})
or one or more
definitions (\eg{} the definitions \verb+(def c "c")+ and \verb+(def d "d")+
in the bottom half of \autoref{fig:move-def-d}).
In the former case, the selected patterns are moved into the
same list pattern in the target position; in the latter case, separate
patterns are kept separate. This is useful for preserving stylistic
choices of the definitions (\eg{} the line length of each definition)
as well as semantic properties (\eg{} dependencies between the
selected definitions).

\parahead{\codeTool{Make Equal by Copying}}

When selecting multiple expressions, this transformation copies one of the
expressions in place of all others.  Unlike the case where all selected
expressions are constant, the key choice here is which expression to use, so the
user is asked to choose which one to replace the others with.

\parahead{\codeTool{Create Function by Merging Definitions}}

While prototyping, it is often convenient to copy-paste code and then
make changes to the different clones as needed. Afterwards, it may be
desirable to pull out the common code between the clones into a single function.
The \codeTool{Create Function by Merging Definitions}
tool takes multiple selected expressions and attempts to abstract them over their syntactic
differences: any differing subtrees become arguments to a new function inserted into the program.
The selected expressions are rewritten as calls to this function.
To avoid suggesting unhelpful small abstractions, the \codeTool{Create Function
by Merging Definitions} tool is displayed only if
the resulting function is larger than a threshold (more precisely, if the number of AST nodes in the function
body is at least double the number of arguments to the function).

\parahead{\codeTool{Create Function from Definition}}

The \codeTool{Create Function from Definition}
tool transforms a selected expression into a lambda abstracted
over constants within the expression body. The transformation provides
two choices: (1) abstracting all constants, or (2) abstracting
constants that are immediately let-bound to variables.
The latter is a (simplified version of a) heuristic proposed by
\citet{sns-uist}. We could add configuration options to ask the user about
all potential constants to abstract; to keep this process lightweight,
however, we propose only the two parameterizations and then allow the user to
modify the result with tools for arguments (below).
After the expression is rewritten, the variable it is
bound to is tracked so that its uses can be rewritten to calls to the
new lambda, with the constants that had been pulled out of the
definition.

\parahead{\codeTool{Add, Remove, or Reorder Arguments}}

The tools to \codeTool{Add, Remove, or Reorder Arguments} allow the interface of a function to be
changed. Arguments may be added to a function by selecting expressions within a lambda and
a target position in the argument list. The \codeTool{Remove} and \codeTool{Reorder Argument} tools allow
the modification to be specified either at the argument list of the function definition or
at a call-site of the function. All three transformations require call-sites to be updated in sync.
Currently, we use a simple
static analysis to track when a lambda is let-bound to a variable.
If the lambda ever escapes this simple syntactic discipline, then we
cannot guarantee that all function calls are rewritten appropriately.
In which case, the transformation is marked as potentially unsafe
(\ie{} yellow). As in other
cases with unsafe transformations, the user must rely on other
mechanisms to ensure correctness (\eg{} types, tests, viewing the output,
or code review).

\parahead{Reorder Items}

This transformation allows one or more (potentially non-consecutive)
items to be removed and inserted elsewhere in the same list. The
whitespace between each pair of consecutive elements is preserved in
the transformed list, a detail that can often be tedious to ensure
with manual text-edits.

\parahead{\codeTool{Rename Variable}}

A transformation commonly found in IDEs
is to \codeTool{Rename} a variable and all its uses. In \deuce{}, the variable to be renamed may
be selected either at
its definition or at one of its usage sites. In either case, as the user
types the new name it is checked to ensure the name will not introduce
collisions; if it would, the transformation is marked as
unsafe (\ie{} yellow) to indicate that a different name may be desired.

\parahead{\codeTool{Swap} Tools}

\codeTool{Swap Variable Names and Usages} can be used to correct the names
chosen for two variables. The alternative is to perform the sequence
of \codeTool{Renaming} the first variable to a temporary, \codeTool{Renaming} the second to
the first, and \codeTool{Renaming} the temporary to the second.
A related transformation, \codeTool{Swap Variables Usages}, is handy for when the
definitions were correct,
but their usages are the opposite of what is intended (\eg{} mixing up
\verb+width+ and \verb+height+ values).

\parahead{\codeTool{Duplicate Definition}}

Text-based copy-paste works especially well when entire, adjacent
lines are copied. For expressions with smaller delimitations (within a
single line) or larger, ``jagged'' ones (different positions across multiple
lines), text-based selection may be more cumbersome.
The \codeTool{Duplicate Definition} transformation is a mouse-based alternative
for copy-pasting an expression to a different target position.

\parahead{\codeTool{Inline Definition}}

This transformation replaces all uses of a selected variable with its
definition. For convenience, multiple different variables can be selected and
inlined simultaneously. As any definition being inlined may itself have
variables, if any such variables are accidentally captured at their new
locations then capture avoiding renamings are offered.

\parahead{\codeTool{Clean Up; Make Single Line; Make Multi-Line; Align Expressions}}

All transformations in \deuce{} attempt to handle whitespace reasonably. Even so,
occasionally a transformation or series of transformations
will result in a program with, for example, a long line of code.
We supplement the whole-program \codeTool{Clean Up} tool of \cite{sns-uist}
with whitespace reformatting rules to break long definitions into
multiple lines and ensure that any multi-line definition is comfortably padded
by a blank line before and after. This is currently the only transformation that
requires no selected items---it applies to the entire program.
We also implement several transformations to help format selected
expressions, by adding or removing line breaks and indentation.

\subsection{Domain-Specific Transformations}
\label{sec:little-transformations-sns}

To demonstrate how our approach can be instantiated with custom
structured transformations, we have implemented several
transformations (in addition to the general-purpose ones)
that are specific to features in \sns{} and/or are particularly
useful for the domain of SVG graphics;
these are summarized in \autoref{fig:transformations-2}.

\parahead{\codeTool{Thaw/Freeze Number; Add/Remove Range; Show/Hide Slider; Flip Boolean}}

Numeric constants in \little{} can be annotated with an optional
range---written \verb+15{1-30}+---to instruct the \sns{}
editor to display a slider in the
output to make it easy to change the number without text-editing.
This feature is an example of ``scrubbing'' constants, a live
programming feature described by Bret Victor
(\url{http://worrydream.com/#!/DrawingDynamicVisualizationsTalk})
and Sean McDirmid
(\url{https://www.youtube.com/watch?v=YLrdhFEAiqo}).
We made one minor addition to the \little{} language: the option to
mark a range annotation as hidden---written
\verb@15{1-30,"hidden"}@---to keep the range information in the
program while suppressing the slider widget in the output.

In \deuce{}, the \codeTool{Add/Remove Range} tool operates on constant literals to attach
or remove these range annotations. New minimum and maximum range
values are determined based on the current value.
The \codeTool{Show/Hide Sliders} tool annotates a
range to be \verb+15{1-30,"hidden"}+, which keeps the range in the
text but suppresses the slider from the output. This tool makes it
quicker to toggle the sliders on and off (and, furthermore, preserves
the possibly-edited min- and max-values to remain in the text).
\sns{} allows numbers to be frozen with the annotation \verb+15!+,
which tells \sns{} not to change this value in response to changes to
the output. (Compared to the discussion of the \codeTool{Create Function from
Definition} tool,
the heuristic for arguments is to choose named \emph{and unfrozen}
constants.)
As with sliders, it can be tedious to use text-editing to
toggle this annotation on and off, so the \codeTool{Thaw/Freeze Constants} tool
provides an alternative.
The \codeTool{Flip Boolean} tool is a quick way to ``scrub'' boolean literals.

\begin{figure*}[b]
\small

\beginToolTable
\toolTableHeaders
\toolCategoryStyle
\rowTool
  {Thaw/Freeze Number}
  {1+}{0}{0}{0}
  {}
\rowTool
  {Add/Remove Range}
  {1+}{0}{0}{0}
  {}
\rowTool
  {Show/Hide Slider}
  {1+}{0}{0}{0}
  {}
\rowTool
  {Rewrite as Offset}
  {1+}{1}{0}{0}
  {}
\rowTool
  {Convert Color String}
  {1+ strings}{0}{0}{0}
  {RGB or Hue}
\hline
\toolCategoryChange
\rowToolNoBreak
  {Flip Boolean}
  {1+ booleans}{0}{0}{0}
  {}
\end{tabular}

\caption{Domain-specific transformations in \deuce{}.}
\label{fig:transformations-2}
\end{figure*}

\parahead{\codeTool{Rewrite as Offset}}

\citet{sns-uist} point out that there are often (at least) three
different common ways to programmatically describe a positional
attribute of a shape: with constants, with constant offsets from an
anchor point, or with constant relative percentages with respect to a
bounding box. When prototyping, it is often easiest to start by
using constants and then later switching to one of the relative versions.
The \codeTool{Rewrite as Offset} tool converts one or more selected constants
into offsets from a selected value. For example, in the expression
\verb+(let x 10 15)+, rewriting \verb+15+ as an offset from \verb+x+
transforms the program to \verb&(let x 10 (+ x 5))&. As with several
other structured transformations we have discussed, this transformation is
conceptually simple but can become tedious and error-prone to perform
with manual text-editing when
multiple numbers scattered across the program need to be offset and when
the base values are not easy-to-remember numbers.

\parahead{\codeTool{Convert Color String}}

In \sns{}, as in HTML and SVG, colors can be specified in a variety of ways: RGBA
codes, HSL codes, HEX, and color strings. The \sns{} editor provides
special support for color attributes defined as ``color numbers''
which are are (essentially) just the hue component of an HSL
triple; in particular, the editor displays a slider next to the shape to
control this color value. The \codeTool{Convert Color String} tool converts
color names (often useful for prototyping) into corresponding numbers, to enable the direct
manipulation support for color numbers that \sns{} provides. This
transformation is an example of how a custom program transformation can
be used to complement other features provided by the IDE.

 \clearpage
\section{Examples in \sns{}}
\label{sec:appendix-examples}

We describe five examples we authored in \sns{}, using a
combination of text- and mouse-based code edits.
Together, the \numExamples{} examples required a
total of \timeRawVideos{} minutes of development time in \deuce{} (the videos
are a few minutes longer because of pauses during the narration), resulting in
approximately \locExamples{} lines of \little{} code for the final programs.
We encourage the reader to watch the accompanying videos, available on the web.
At several points in the videos, we use the existing \sns{} ability to directly
manipulate the size, position, and color of shapes in the output, as a way to
indirectly update constant literals in the program. We use this feature for
simplicity; the constant literals could, of course, be changed with ordinary
text-edits.

\parahead{Preliminary Version of \deuce{}}

Note that the screenshots below, as well as the videos, show a preliminary
version of \deuce{} (implemented in \sns{}~\version{0.6.0}),
with cosmetic differences compared to the bounding boxes
and target positions shown in this paper. In the preliminary version,
\codeTool{Create Definition from Definition} was called \codeTool{Abstract},
\codeTool{Create Definition by Merging Expressions} was called \codeTool{Merge
Expressions}, and the two \codeTool{Make Equal} tools were combined into a
single tool.


\subsection{Example: \exampleZero{}}

We use an example in this section to motivate and summarize the
workflow enabled by \deuce{}. The task, to write a program that
generates an SVG design, is borrowed from \citet{sns-uist}.
Throughout the discussion, notice how \deuce{} automates tasks
that would otherwise be tedious and error-prone, and the user
performs manual text-edits for tasks that require additional human
insight and choice.

\setlength{\intextsep}{6pt}%
\setlength{\columnsep}{10pt}%
\begin{wrapfigure}{r}{0pt}
\includegraphics[scale=0.20]{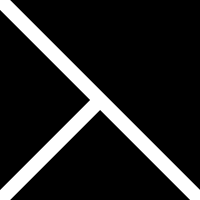}
\end{wrapfigure}
Consider the task of writing a program that generates the \sns{} logo
(shown in the adjacent screenshot), where two white lines
atop a black rectangle reveal a lambda symbol between three triangles.
We would like the program to be structured so that it can be reused
to easily generate configurations of the logo with different design parameters,
namely, the size, background color, and color and width of the
lines. We will describe a sequence of text- and mouse-based code
edits in \deuce{} that allows the user to
prototype, repair, and refactor the code until it achieves
the desired goal. The reader is encouraged to follow along with
the example in our online demo, or to watch the
video.

\parahead{Phase I: Prototyping}

\begin{figure}[h]
\centering
\includegraphics[scale=\snsScreenshotScale]{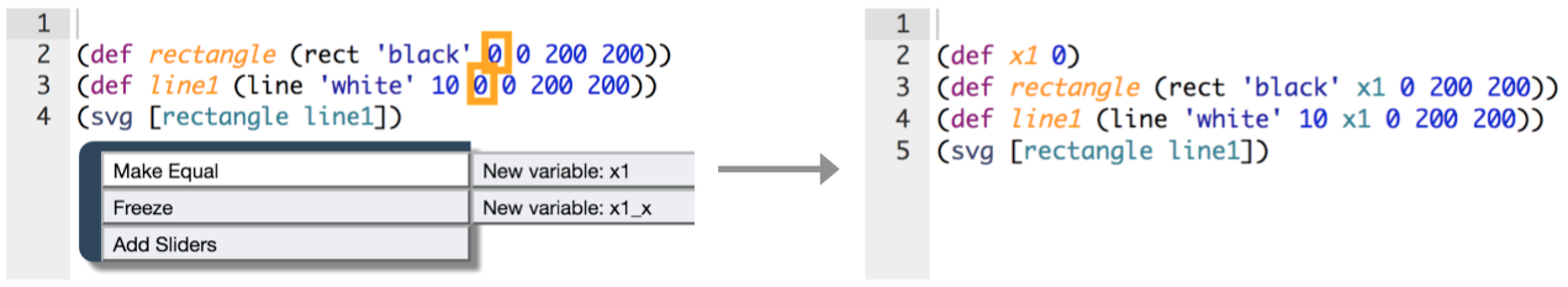}
\caption{When the user holds the Shift key and selects the two constants
(highlighted in orange), a context-sensitive menu identifies the program
transformations that may be applied. When the user invokes the ``Make
Equal, New variable: x1'' tool, the program is transformed to the
version on the right.}
\label{fig:overview-1}
\end{figure}

The user writes definitions for the background rectangle and the first line,
between the top-left and bottom-right corners of the logo.
To start, these shapes are both
positioned at the origin \verb+(0,0)+ and stretch to
\verb+(200,200)+. They are combined in
\verb+(svg [rectangle line1])+ to generate the first version to
render; this code can be seen in the screenshot in the left half
of \autoref{fig:overview-1}.
The user confirms that the pieces look roughly as intended, and
now returns to the program to replace the hard-coded \verb+(0,0)+
constants with variables so that the positions of both shapes can be
modified together.

Rather than text-editing, however, the
user can use the structured editing capabilities of \deuce{} to
perform this task in a quick, and safe, way.
The user holds down the Shift key, which causes \deuce{} to display
clickable widgets when hovering over different parts of the program
text. The user clicks on the two \verb+0+ constants that correspond to
the x-positions of the top-left corners of the shapes, and \deuce{}
displays a menu of potential transformations underneath the selected
widgets. When the user hovers over the \codeTool{Make Equal} tool in the menu,
\deuce{} displays a list of candidate new variable names to add to the
program. When hovering over each choice, \deuce{} previews the result
of the transformation; in each case, the selected constants are replaced by
uses of the new variable, which binds \verb+0+ and is defined at the
innermost scope relative to the constants.
The user chooses the new variable name to be \verb+x1+, and
the code is transformed as shown in the right half of
\autoref{fig:overview-1}.

\begin{wrapfigure}{r}{0pt}
\includegraphics[scale=\snsScreenshotScale]{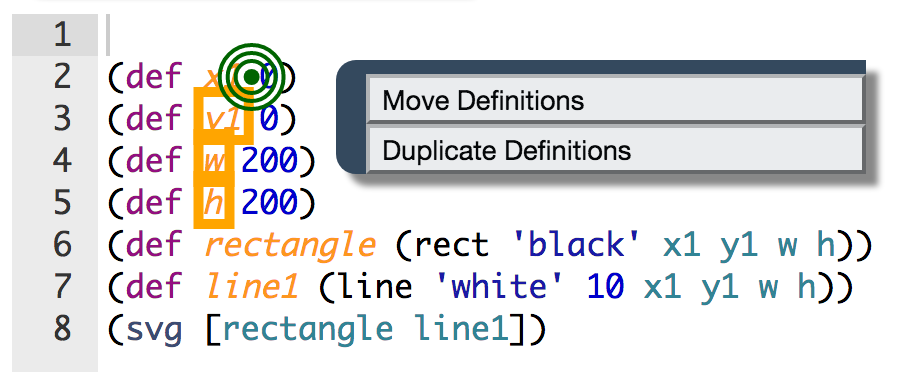}
\end{wrapfigure}

Next, the user employs the \codeTool{Make Equal} tool three more times, selecting the
remaining pairs of constants that define the x- and y-positions of the
top-left and bottom-right corners. Although happy with the four
names \deuce{} has suggested (\verb+x1+, \verb+y1+, \verb+w+, and
\verb+h+), the user wants to combine the
definitions into a single line, because each of the names and values is
short.
The user selects three variables to move---\verb+y1+,
\verb+w+, and \verb+h+---and the \emph{target position} after the
\verb+x1+ variable, to indicate that the selected variables should
appear inline after the \verb+x1+ (shown on the right).
The user hovers over the \codeTool{Move
Definitions} tool to preview the transformation where all four
variables are defined in a tuple (\ie{} 4-element list), and then
selects this transformation.

Now, the user uses normal copy-paste to duplicate the \verb+line1+ definition,
renames it \verb+line2+, and adds \verb+line2+ to the list of shapes.
This second line will eventually connect the bottom-left corner of the
logo to its center, which will form the lambda symbol.
To start testing, however,
the user edits the start and end points to be \verb+(x1,h)+ and
\verb+(w,y1)+, respectively, which generates the symbol ``X'' when run.

\begin{wrapfigure}{r}{0pt}
\includegraphics[scale=\snsScreenshotScale]{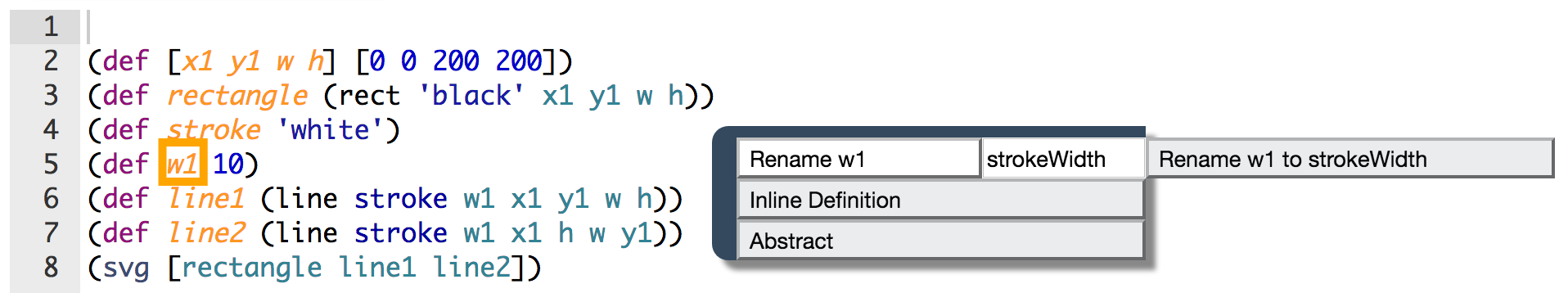}
\end{wrapfigure}

The user invokes
\codeTool{Make Equal} twice to relate the stroke color and width of the two
lines with two new variables called \verb+stroke+ and \verb+w1+,
respectively.
The user is not happy with the name chosen for the latter, so
she clicks on the variable definition and uses the \codeTool{Rename} tool to rename
\verb+w1+ to \verb+strokeWidth+ (the interaction is shown on the
right), which automatically replaces all uses with the new name.

\begin{wrapfigure}{r}{0pt}
\includegraphics[scale=\snsScreenshotScale]{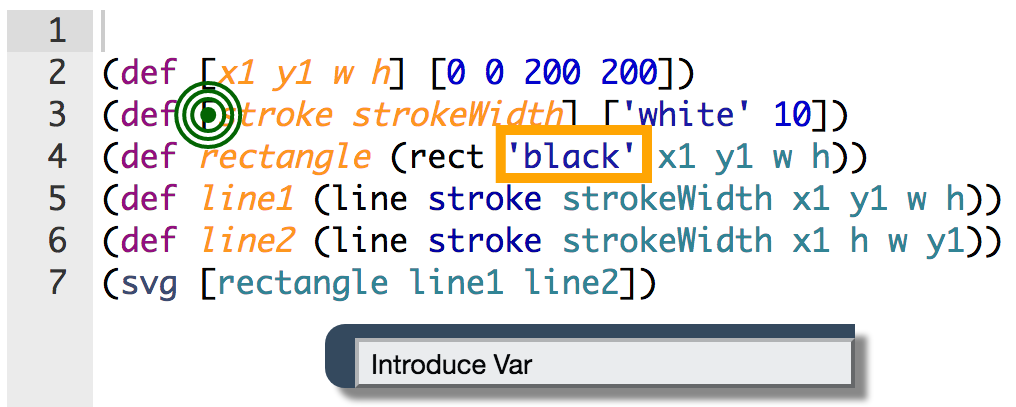}
\end{wrapfigure}

Then the user invokes the \codeTool{Move Definition} tool to combine these two
variables into a tuple (not shown). The user invokes \codeTool{Move Definition}
once again
to move \verb+rectangle+ below \verb+stroke+ and \verb+strokeWidth+ (not shown).
Having taken care to organize the code in a readable fashion, the
user would like to define a variable to clearly identify that the
constant \verb+'black'+ defines the color of the rectangle. The user
selects this constant and the target position before \verb+stroke+,
and invokes the \codeTool{Introduce Variable} tool (shown on the right) to add a
new variable called \verb+fill+ in place of the string literal.

\parahead{Phase II: Repairing}

At this stage, the user has become comfortable with the basics of the
design, but is aware of two issues that must be addressed.
The first issue is that even though the position of the top-left
corner has been factored into variables \verb+x1+ and \verb+y1+,
the relationships for the other endpoints depend on the values of these
variables both being \verb+0+. To verify this,
the user text-edits them to be \verb+50+ and \verb+50+, re-runs the
program, and confirms that the lines do not ``move'' with the rectangle.
Knowing what the intended relationships ought to be, the user
text-edits the second endpoint of \verb+line1+ to be \verb@(+ x1 w)@ and
\verb@(+ y1 h)@.

\begin{wrapfigure}{r}{0pt}
\includegraphics[scale=\snsScreenshotScale]{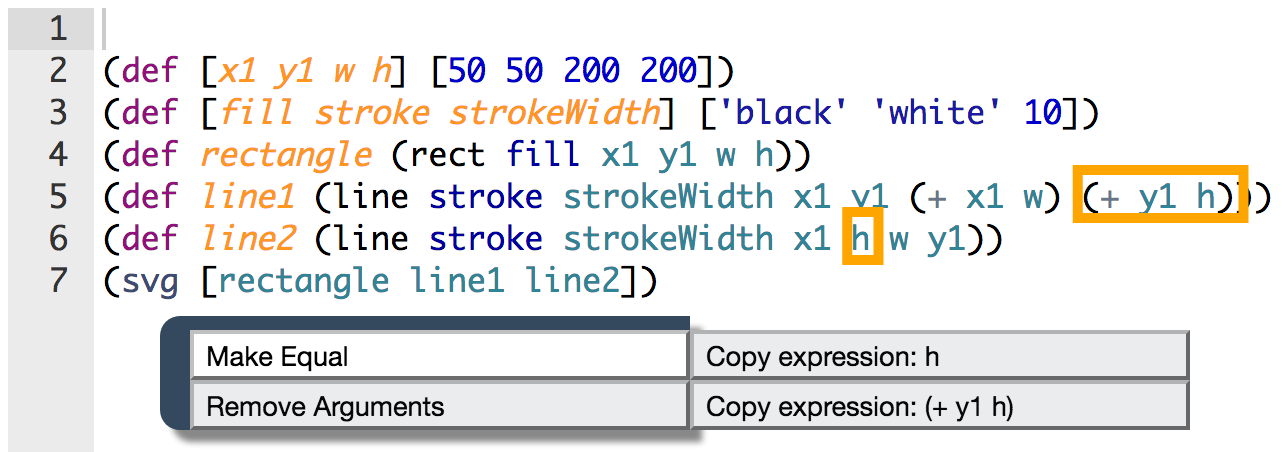}
\end{wrapfigure}

Now, to snap the other line of the ``X'' to the
top-right corner, the user must use these same subexpressions in the
definition of \verb+line2+. The user selects \verb@(+ y1 h)@ of \verb+line1+
and \verb+h+ in \verb+line2+, and invokes the \codeTool{Make Equal} tool (shown
on the right). Because these
two subexpressions differ, the \deuce{} menu asks the user to select which
of these expressions should be used in both places; the user chooses
to use the sum expression. Similarly, the user invokes \codeTool{Make Equal} (not
shown) to replace the \verb+w+ in \verb+line2+ with the \verb@(+ x1 w)@.

The remaining issue is that the second line needs to be half the
length, so that it reveals the lambda symbol instead of the letter
``X.'' To do this, the user text-edits the coordinates of the
endpoint to be \verb@(+ x1 (/ w 2))@ and \verb@(+ y1 (/ h 2))@,
respectively. Viewing the output now reveals the lambda. Either with
text-edits or the existing output-directed manipulation features of
\sns{}, the user
varies the values of the four positional variables, and visually
confirms that the output continues to exhibit the intended lambda
symbol.

\parahead{Phase III: Refactoring}

At this point, the user has finished encoding all the desired
relationships in the program. Now is the time to refactor the program
so that it can be reused to generate multiple variations. First, the
user selects the list of shapes at the end of the program
and invokes the \codeTool{Introduce Variable} tool
(shown below left) to give it a name (\verb+shapes+) outside the \verb+svg+
expression. Next, the user selects the definitions that contribute to
\verb+shapes+, and invokes the \codeTool{Move Definitions} tool to place them inside
the \verb+shapes+ definition (shown below right).

\begin{figure}[h]
\centering
\begin{minipage}{0.49\textwidth}
\centering
\includegraphics[scale=\snsScreenshotScale]{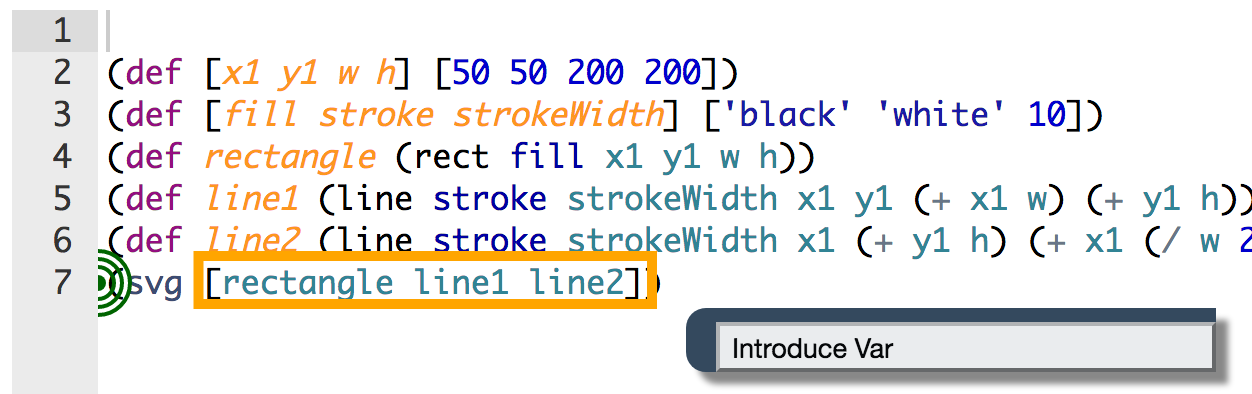}
\end{minipage}
\hspace{0.0\textwidth}
\begin{minipage}{0.49\textwidth}
\centering
\includegraphics[scale=\snsScreenshotScale]{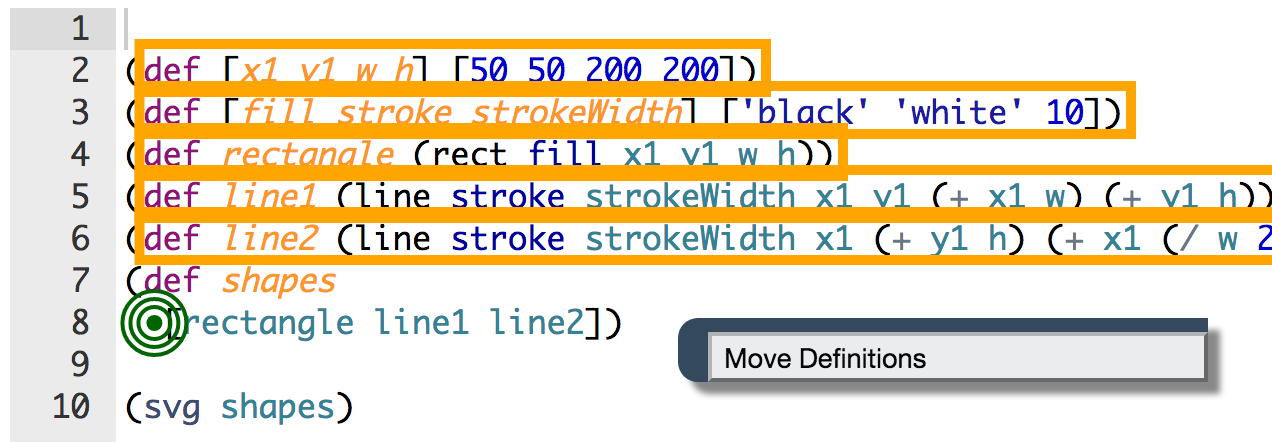}
\end{minipage}
\end{figure}

\noindent
The top-level definitions are turned into local
let-bindings, taking care of indentation and parenthesis delimiters
that would otherwise require tedious, manual text-edits. The user uses
\codeTool{Rename Variable} (not shown) to change \verb+shapes+ to \verb+logo+.

\begin{figure*}[h] 
\includegraphics[scale=0.50]{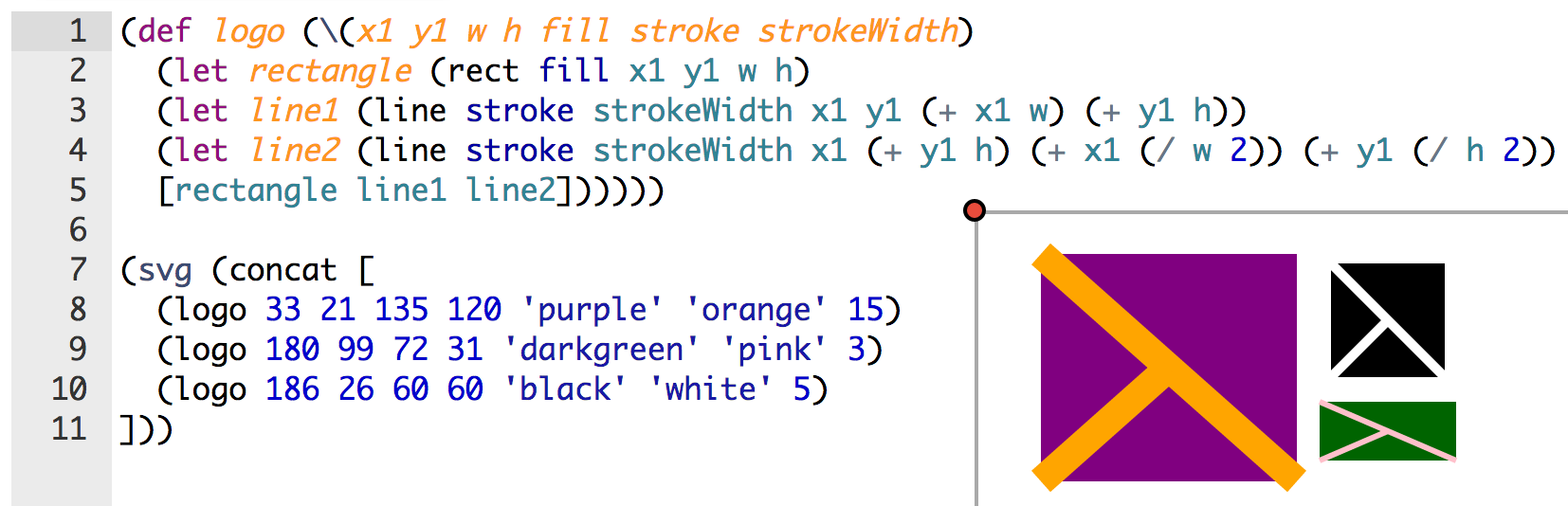}
\caption{Program to generate \sns{} logo,
developed with a combination of text-edits and \deuce{} code tools.}
\label{fig:overview-variations}
\end{figure*}

The final step
is to turn \verb+logo+ into a function that is parameterized over the
design constants inside the definition. Selecting the definition (shown below),
\deuce{} shows a \codeTool{Abstract} tool to turn several of the constants into
function arguments.
In \autoref{fig:overview-variations},
notice how the use of \verb+logo+ has been rewritten to a call, with
the appropriate arguments selected from their values within the
definition. Again, this would be a tedious and error-prone manual
transformation, as the connection between formal parameters and
actual arguments are
not syntactically adjacent in the program. However, this
transformation is easy to automate with structure information.
To create other versions, the user copies and pastes the
function call and changes the arguments to each (shown in
\autoref{fig:overview-variations}).

\begin{wrapfigure}{r}{0pt}
\includegraphics[scale=\snsScreenshotScale]{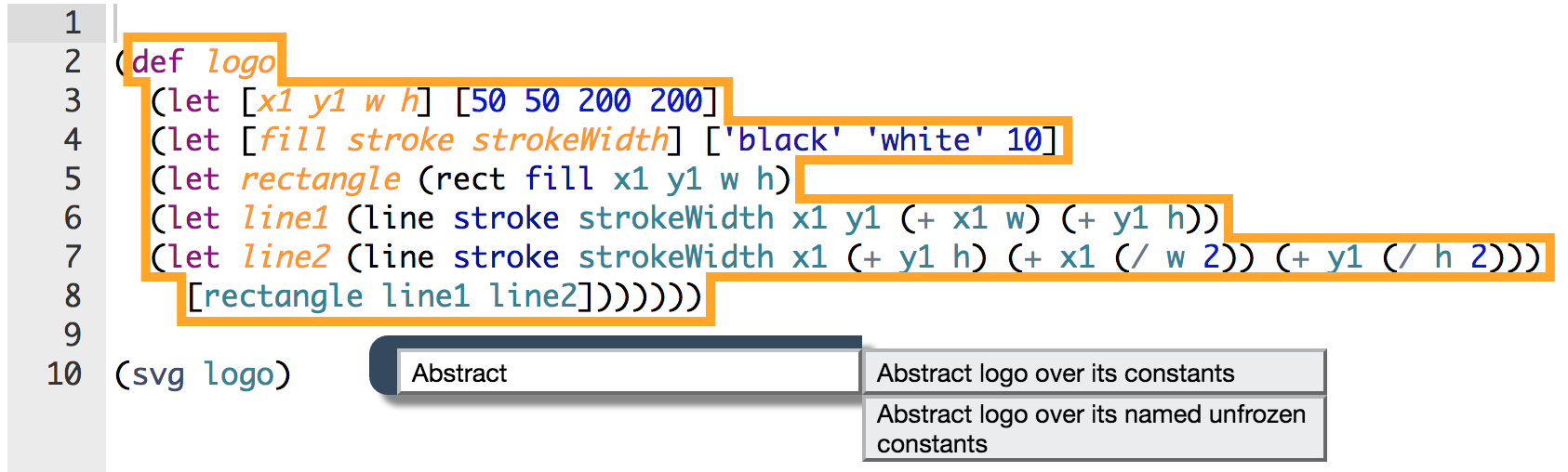}
\end{wrapfigure}

To recap, the process of prototyping, repairing, and refactoring the
program is a text-driven process, as usual, but with support for
automating low-level, structural edits that are tedious and
error-prone (\eg{} because of typos, accidental shadowing, and mismatched
delimiters). Furthermore, the tools can suggest useful information,
such as the different variable names offered by Make Equal and
Introduce Variable. As a result, the user spends keystrokes on the more
interesting tasks that are harder to automate---arithmetic
relationships in the design,
the choice of names, and
final decisions about whitespace and formatting.

\subsection{Example: \exampleOne{}}

\begin{wrapfigure}{r}{0pt}
\includegraphics[scale=0.10]{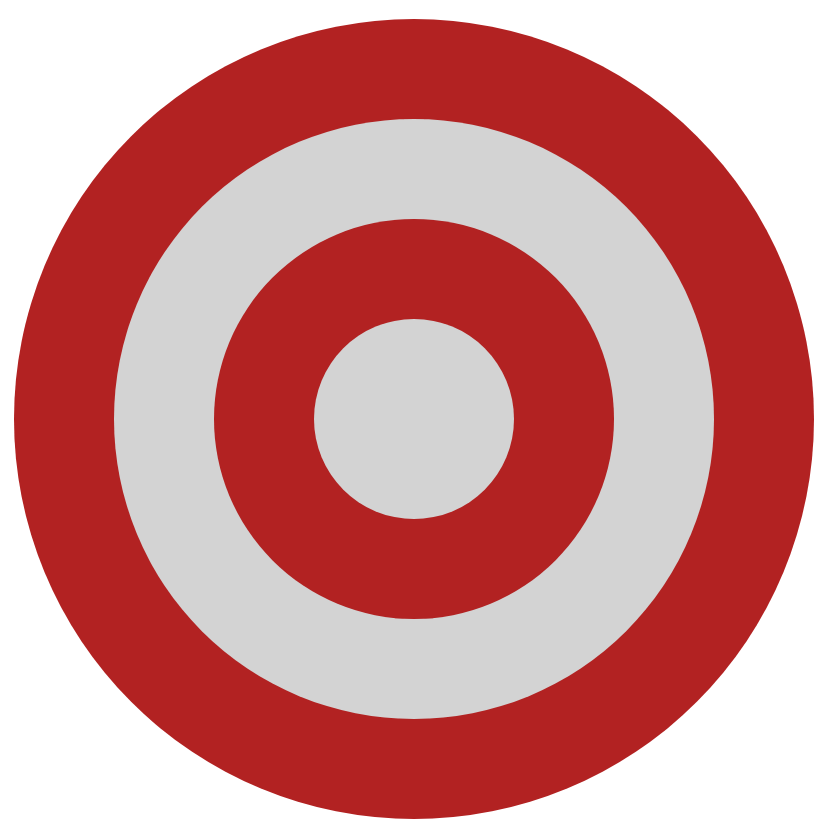}
\end{wrapfigure}

In this example, our goal is to generate a target comprising
concentric rings of alternating color. We start by writing a single
red circle, using the expression \verb+(* 1 50)+ for the radius to
anticipate that it will scale linearly for ring \verb+2+, \verb+3+, and
so on. We use \codeTool{Abstract} to extract a function,
\verb+ring+, that is parameterized over this index \verb+i+, and we use
\codeTool{Remove Arguments} so that \verb+ring+ is parameterized over
only \verb+i+. We use \codeTool{Introduce Variable} to give a name to the ring color,
and then use text-editing to choose red or gray depending on the
parity of index \verb+i+.
Next, we update the main expression with text-editing to \verb+map+
the function's \verb+i+ over the list of indices \verb+(reverse (range 1 4))+.
We use the \codeTool{Abstract} tool to extract a \verb+target+ function
for drawing these concentric circles. We would like the
\verb+target+ function to take \verb+cx+, \verb+cy+, \verb+ringWidth+, and
\verb+ringCount+, but the first three of these are currently constants
inside the \verb+ring+ function. To turn these constants into function
arguments, we move the entire \verb+ring+ definition (and thus the constants of interest)
inside of \verb+target+, which then allows us to use the \deuce{} tools to
introduce and rename the desired arguments.

\subsection{Example: \exampleTwo{}}

\begin{wrapfigure}{r}{0pt}
\includegraphics[scale=0.30]{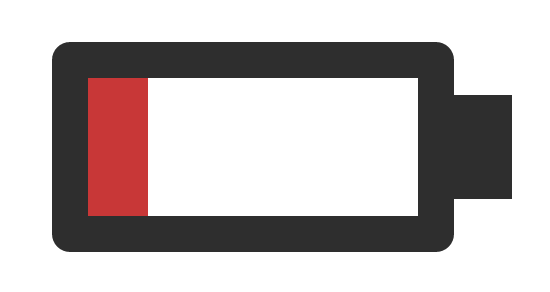}
\end{wrapfigure}

In this example, our goal is to build a program that generates a
battery icon, akin to those often found in operating system task bars.
The design comprises three shapes: a polygon with rounded stroke
for the body of the battery, a rectangle for the cap, and a
rectangle for the battery juice inside. In addition to setting up the
appropriate positional relationships, we want the color of the battery
juice to change based on the amount that remains.
Our development process from start to finish, which takes
approximately \timeRawVideoBattery{} minutes (without narration),
mixes text-edits and \deuce{} transformations throughout. Our
general workflow is to incrementally add new shapes and features,
often starting with hard-coded or copy-pasted expressions, and then
iteratively repairing the program by adding new variables and
relationships.

The first shape we add to the program is the polygon for the body outline. We use the
\codeTool{Introduce Variable} tool to give names to the top-left and top-right
corners of the body, which are needed by subsequent expressions.
We use the \codeTool{Make Equal} tool to equate certain offsets among edges in
the design, and we use the domain-specific \codeTool{Freeze} tool to ensure that
some of these offsets are always the constant zero. If we accidentally
swap the usages of \verb+width+ and \verb+height+ variables,
we can use the \codeTool{Swap Variable Names and Usages}
tool to correct the bug. The second shape we add to the program
is the cap. Again, we use a combination of text-edits and structured
editing tools, such as \codeTool{Introduce Variable} and \codeTool{Move Definitions},
during this process.

The last shape we add is the colored rectangle for the battery juice.
After we add it to our list of shapes, we see that the juice appears
on top of the body rather than behind it. We use the \codeTool{Reorder Items}
tool to move this rectangle earlier in this list, which results in the
desired z-ordering.
When prototyping, it is
natural to define the width of this rectangle directly as a constant
\verb+w+, with the intention that it remains less than the \verb+width+ of the body.
Later in the development process,
we use text-editing to introduce a conditional
expression that determines the color (\ie{}, green, black, orange, or
red) based on the value of the percentage \verb+(/ w width)+. This
expression appears in several guards of a multi-way conditional, so
we use the \codeTool{Make Equal with Single Variable} tool to give it a name,
which we \codeTool{Rename} to \verb+juicePct+. Now that the relationships are set
up, we realize that it would be better to first define \verb+juicePct+
(the percentage denoted by a number between \verb+0.0+ and \verb+1.0+)
and then derive \verb+w+ in terms of it. We use the \codeTool{Move Definition}
tool to drag the former above the latter, and \deuce{} proposes an
option where the definitions are inverted, specifically,
\verb+juicePct+ is redefined to be a constant percentage and \verb+w+
is defined in terms of it. We use the \codeTool{Add Range} tool twice twice, once
on \verb+w+ before it was rewritten and once on \verb+juicePct+
afterwards. In each case, the automatically chosen range was useful
for allowing the slider to quickly manipulate reasonable values for
each quantity.

At this point, our program generates the entire icon in terms of the design
parameters. To finish, we turn the definition into a function
using a similar series of \codeTool{Move Definitions} and \codeTool{Create
Function from Definition} transformations as described earlier.
This time, we realize
that the \codeTool{Abstract} tool did not make all of our desired constants into
parameters. So, we use the \codeTool{Add Argument} and \codeTool{Rename} tools to reach our
desired parameterization of \verb+topLeft+, \verb+bodyWidth+, \verb+bodyHeight+,
\verb+capWidth+, \verb+capHeight+, \verb+strokeWidth+, and \verb+juicePct+.

\subsection{Example: \exampleThree{}}

In this example, our goal is to build a program that generates a
coffee mug, in a way that it is easy to reposition and resize the logo.
When developing the body of the mug and two concentric ellipses for
the handle, the \codeTool{Introduce Variable} and \codeTool{Move Definition} tools help turn
initially hard-coded shapes into the desired relationships.

\begin{wrapfigure}{r}{0pt}
\includegraphics[scale=0.22]{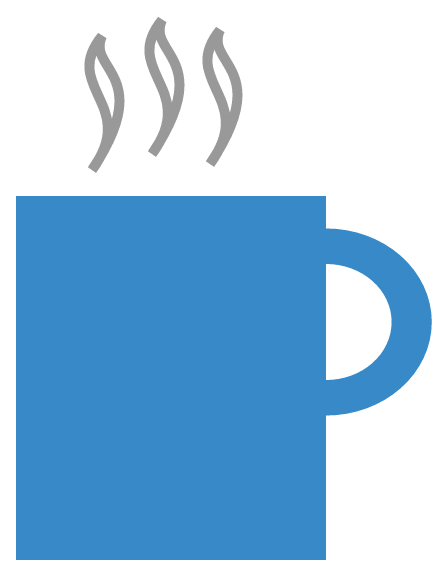}
\end{wrapfigure}

When designing the steam, we use tools not already exercised in the previous
examples. When we add the first steam puff, we use hard-coded
constants for all of the points and control points of the path. This
helps us get the initial design for the curvy puff, but makes it hard
to move to a different position; all 12 constants must be updated by
the same offset to translate it. We use the \codeTool{Rewrite as Offset} tool
several times to make the steam puff rigid.
Then we use the \codeTool{Duplicate Definition} tool to copy-paste (via \deuce{}
rather than text-editing) the first steam definition twice. After
changing just the initial position of each puff, our copy-pasted
definitions contain nearly identical code. We use the \codeTool{Merge} tool to abstract the
three steam puffs over their differences (\ie{} the position).
We use \codeTool{Rewrite as Offset} several more times to position the
second and third steam puffs in terms of the first, and then again to
position the first steam puff in terms of the mug; the effect is that
the steam remains rigid and correctly positioned as the mug is
translated to different positions. During these last steps, we move
several definitions from the bottom of the program up to the top so
that the related expressions are closer together; the \codeTool{Move Definition}
tool allows us to make such transformations without fear of breaking
dependencies in or changing the binding structure of the program.

\subsection{Example: \exampleFour{}}

\begin{wrapfigure}{r}{0pt}
\includegraphics[scale=0.17]{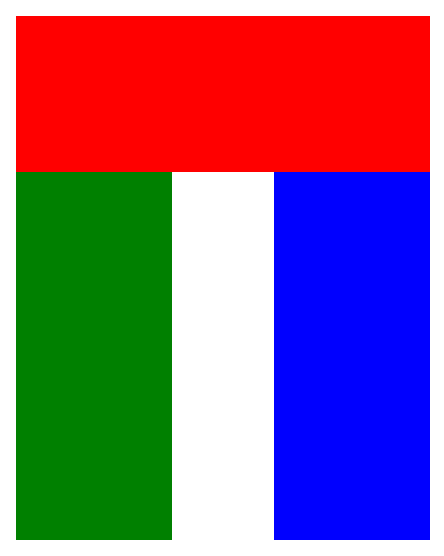}
\end{wrapfigure}

Inspired by the Mondrian programming-by-demonstration graphics editor~\cite{Mondrian},
our final example is an arch, where two upright
rectangles support a third horizontal rectangle, all of which are of
equal width. As in the previous examples, we use tools like \codeTool{Introduce
Variable} often to help reorganize the code and text-edits to fill in
arithmetic relationships. Unlike the previous examples, we use tools
that manipulate concrete whitespace---\codeTool{Make Multi-Line} to facilitate
the step of going from one call to \verb+rect+ to multiple ones, and
\codeTool{Make Multi-Line} to make the arguments to these adjacent calls line up
vertically, making it easier to distinguish the differences between
all calls. These tools eliminate some of the tedious text-edits that
arise when making such stylistic changes to the code.

 \clearpage
\section{User Study}
\label{sec:appendix-user-study}

We configured a pared down version of the system that turned off all \sns{}
features unrelated to the interactions being studied. This version also
contained a panel in the bottom-right corner with instructions about the current
task. To improve readability, the screenshots below show a lighter color theme
than used in the user study)
The user study version of the system, as well as a video that
demonstrates how to complete the following tasks, is available
at \url{http://ravichugh.github.io/sketch-n-sketch/blog/}.

\subsection{Tasks}

\newcommand{\taskScreenshotScale}{0.33}

\parahead{Head-to-Head Task: One Rectangle}

This task required swapping two arguments to a function (either with
Swap Expressions or Swap Variable Usages), and combining and reordering
five separate definitions into a single tuple definitions (requiring at least
two calls to Move Definitions).

\begin{figure}[h]
\center{\includegraphics[scale=\taskScreenshotScale]{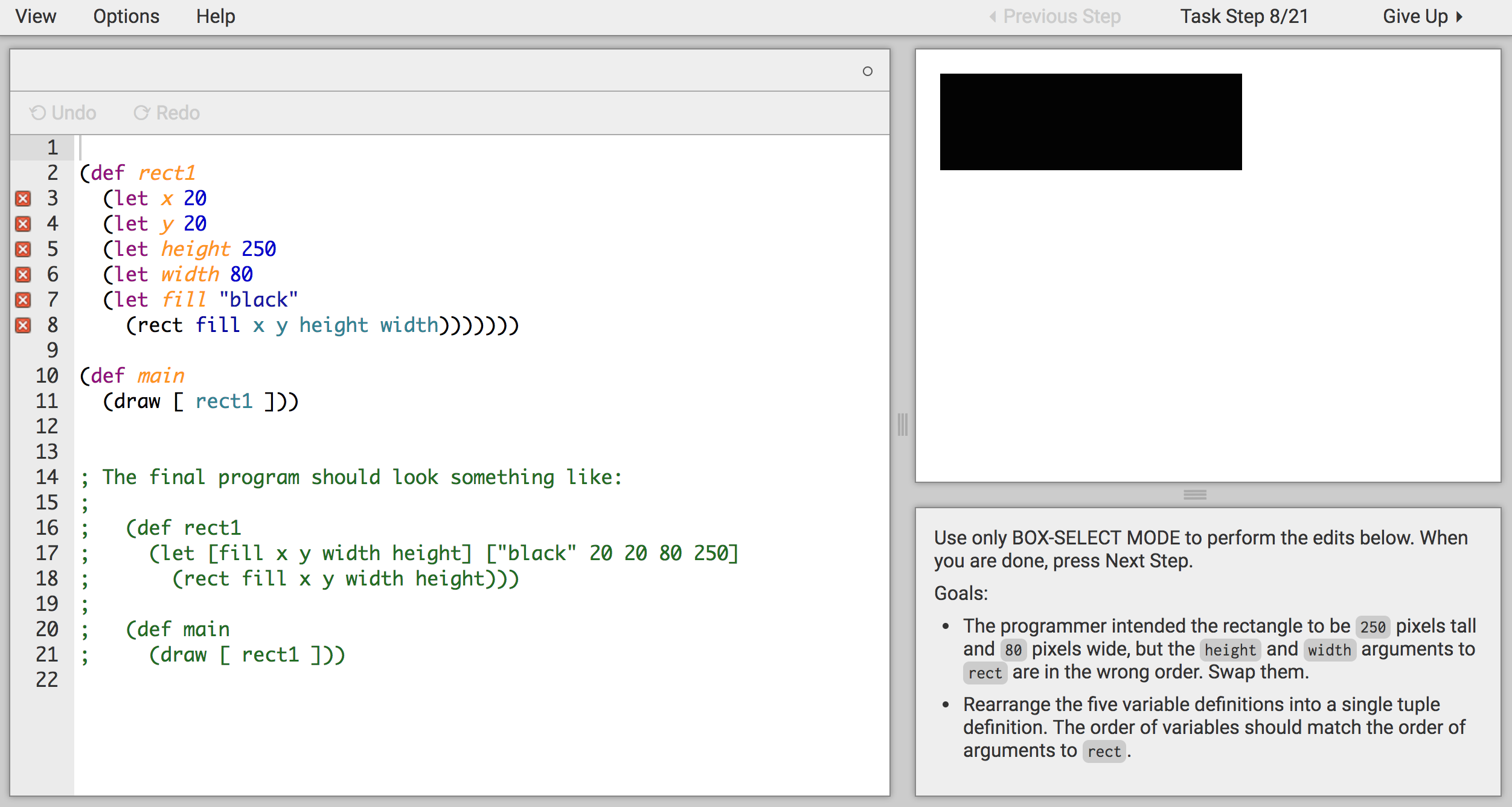}}
\end{figure}

\parahead{Head-to-Head Task: Two Circles}

This task (described as Example 1 in \autoref{sec:overview}) required turning
a definition into a function (Create Function from Definition or Create Function
from Arguments) and then rearranging arguments (Reorder Arguments).

\begin{figure}[h]
\center{\includegraphics[scale=\taskScreenshotScale]{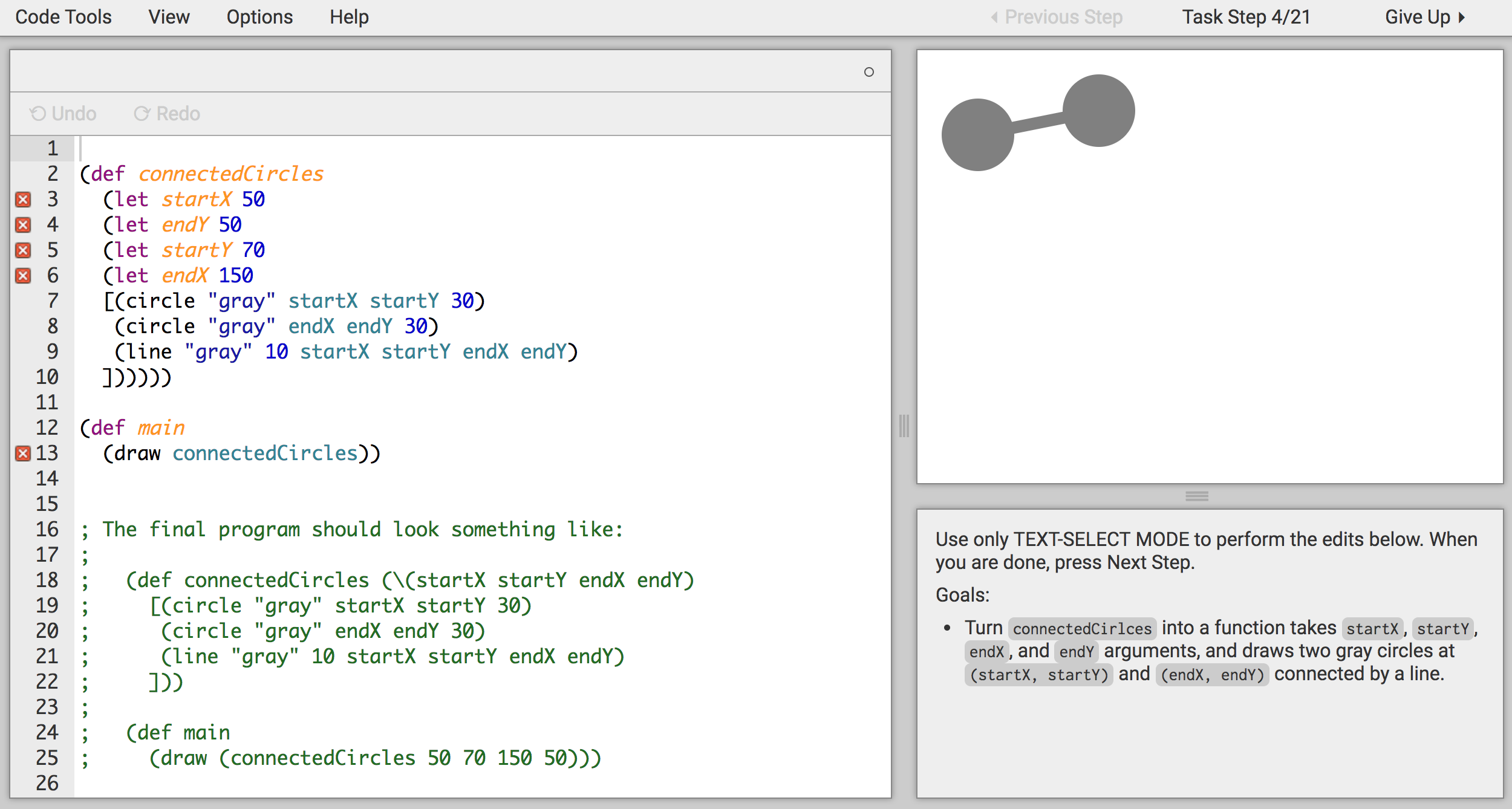}}
\end{figure}

\clearpage

\parahead{Head-to-Head Task: Three Rectangles}

This task required factoring three nearly-identical definitions into a helper
function (Create Function by Merging Definitions) and renaming the resulting
function (Rename Variable).

\begin{figure}[h]
\center{\includegraphics[scale=\taskScreenshotScale]{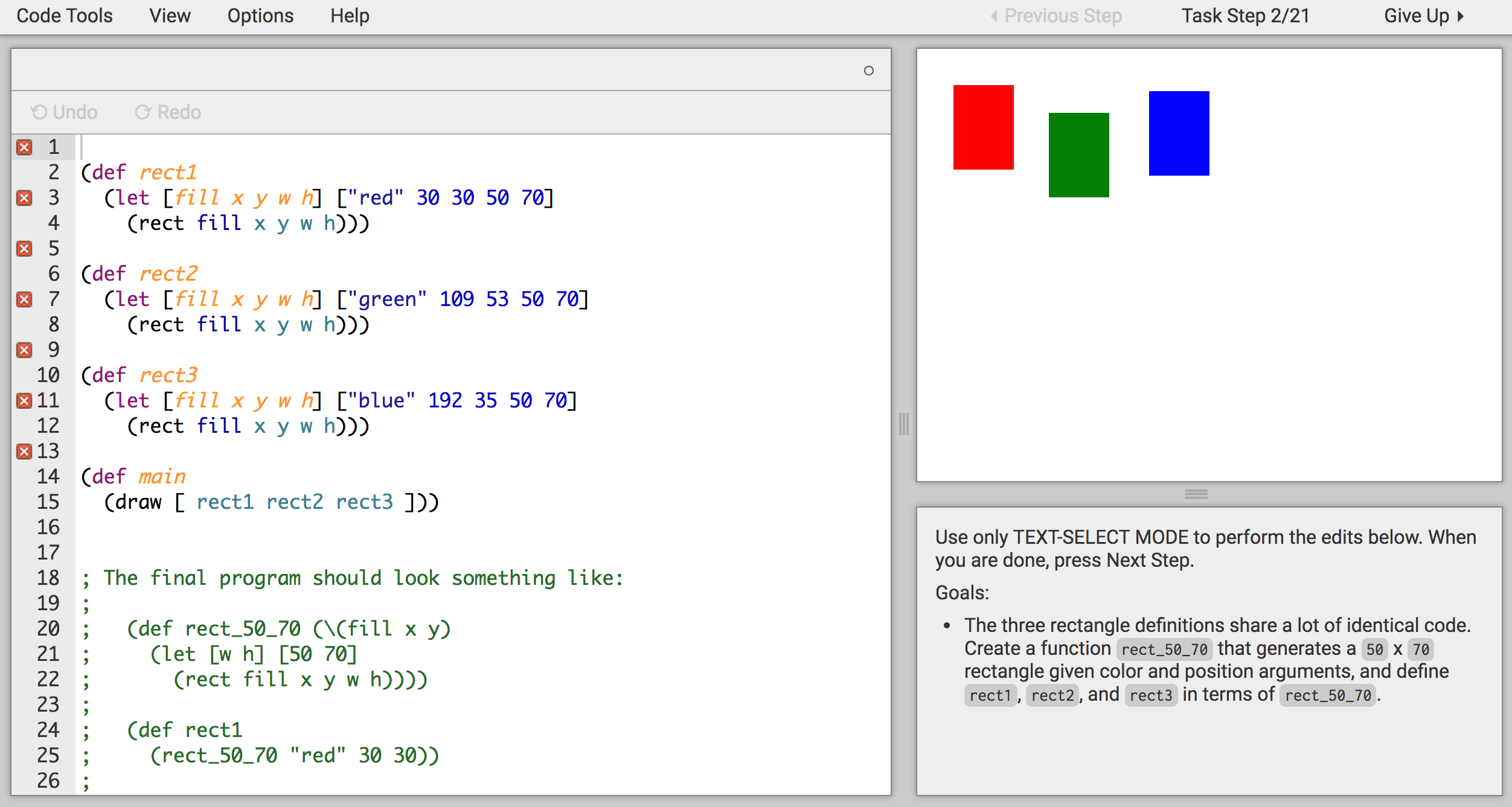}}
\end{figure}

\parahead{Head-to-Head Task: Four Circles (a.k.a. Target Icon)}

This task required removing a function argument (Remove Argument), renaming a
function argument (Rename Variable), moving a function definition inside another
(Move Definition), and adding several additional arguments to an existing
function (Add Arguments).

\begin{figure}[h]
\center{\includegraphics[scale=\taskScreenshotScale]{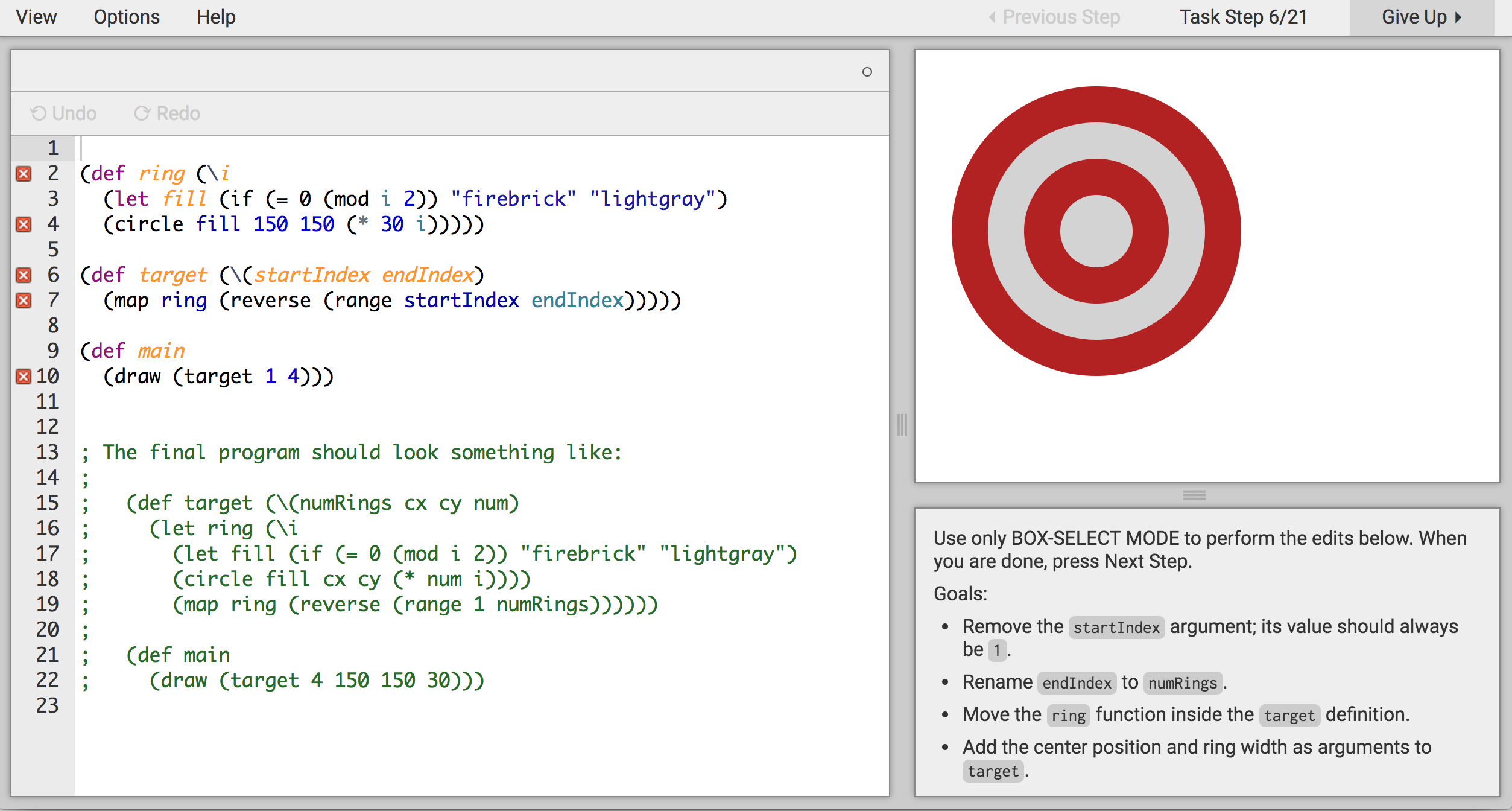}}
\end{figure}

\clearpage

\parahead{Open-Ended Task: Four Squares}

This task required factoring four similar function calls into a helper function
(Create Function by Merging Definitions), 
creating a function (Create Function from Definition followed by Add Arguments,
or Create Function from Arguments), and renaming five variables (five uses of
Rename Variable).

\begin{figure}[h]
\center{\includegraphics[scale=\taskScreenshotScale]{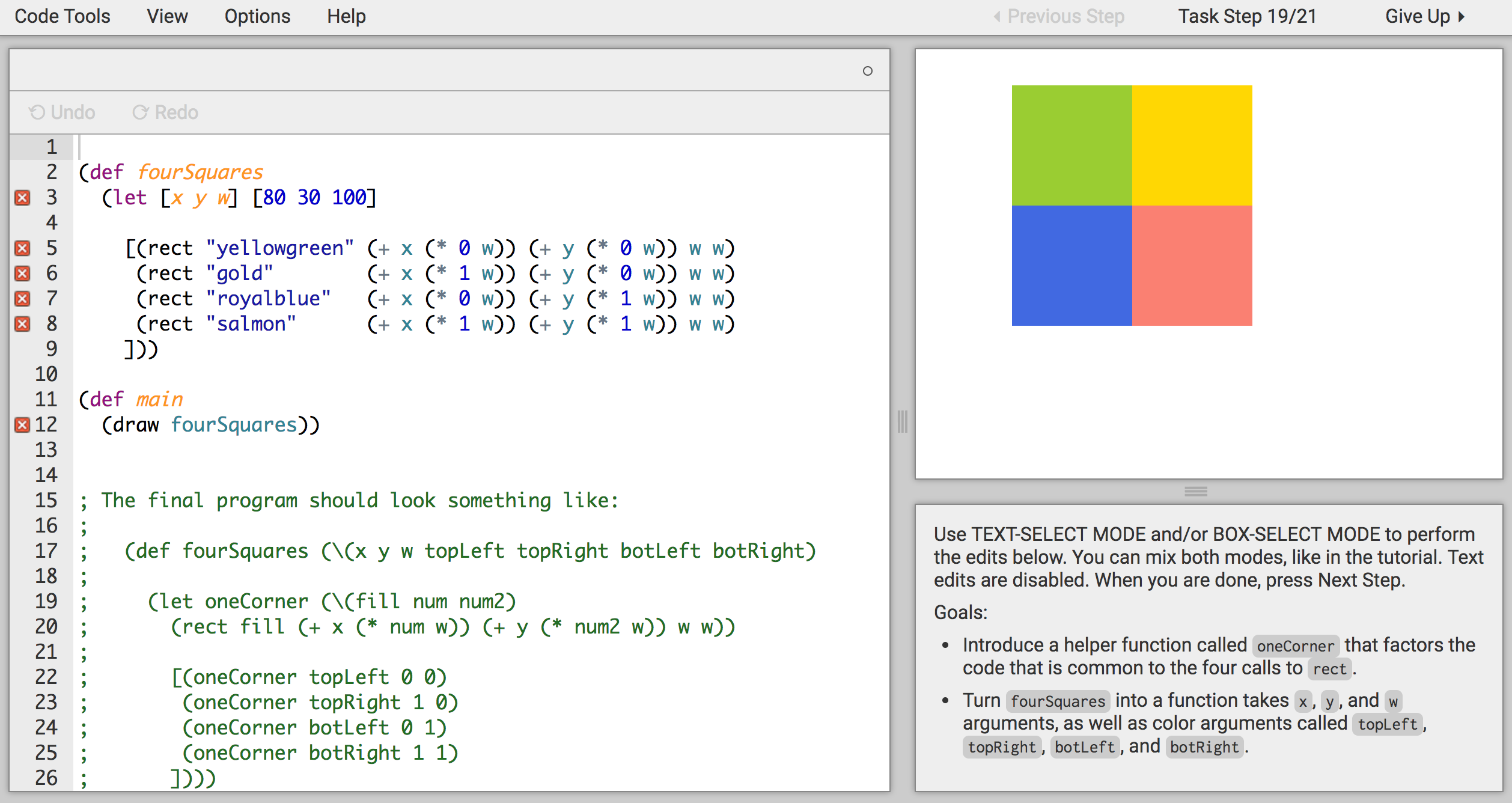}}
\end{figure}

\parahead{Open-Ended Task: Lambda Icon}

This task is the same as Example 3 in \autoref{sec:overview}, not including the
last step to Move Definitions. That is, there are seven variables to define
(one using Introduce Variable and six with Make Equal with Single Variable).
These seven variables must be defined in two tuples, which can be accomplished
by Move Definitions but also without it if the previous tools were used with the
appropriate optional target positions.

\begin{figure}[h]
\center{\includegraphics[scale=\taskScreenshotScale]{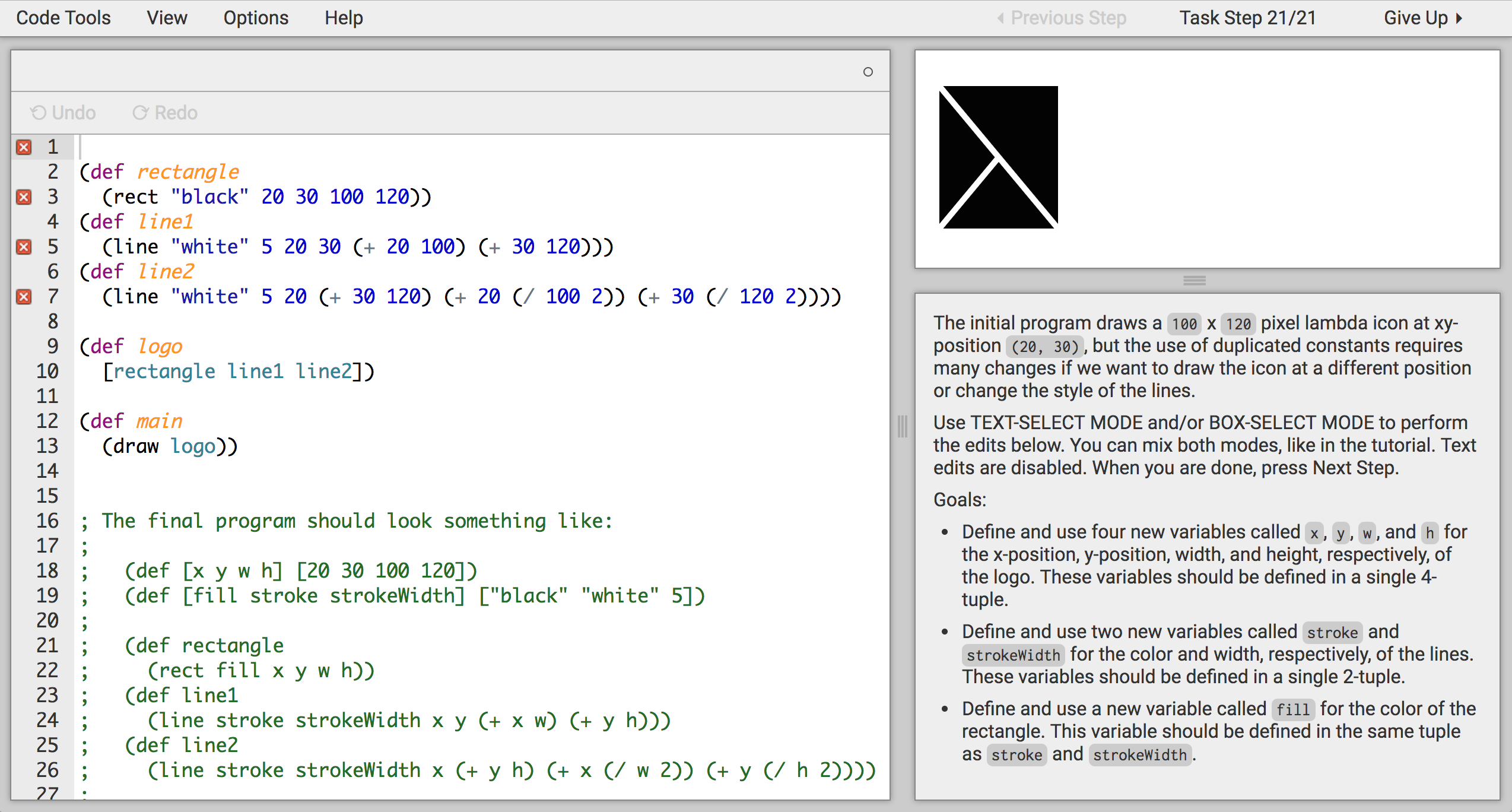}}
\end{figure}

\clearpage

\subsection{Exit Survey}

\setlength{\parindent}{0em}
\setlength{\parskip}{1em}


\subsection*{Background Questions}

How many years of programming experience do you have?

\begin{itemize}
  \item Less than 1
  \item 1-2
  \item 3-5
  \item 6-10
  \item 11-20
  \item More than 20
\end{itemize}

How many years of functional programming experience do you have
(in languages such as Racket, Haskell, OCaml, Standard ML, or Elm)?

\begin{itemize}
  \item Less than 1
  \item 1-2
  \item 3-5
  \item 6-10
  \item 11-20
  \item More than 20
\end{itemize}

Do you use languages and tools that provide automated support
for refactorings or other program transformations?
(Examples include Java and Eclipse, Idris and its editor, etc.)
If so, please describe which tools and how often you use them.

Did have knowledge or hands-on experience with Sketch-n-Sketch
before this study?

\begin{itemize}
  \item Yes
  \item No
\end{itemize}

Did have knowledge or hands-on experience with the
Code Tools in Sketch-n-Sketch before this study?

\begin{itemize}
  \item Yes
  \item No
\end{itemize}

If you answered Yes to either question above,
please explain briefly.

\clearpage

\subsection*{Head-to-Head Comparisons on First Four Tasks}

\newcommand{\interactionA}{Text-Select Mode}
\newcommand{\interactionB}{Box-Select Mode}

\newcommand{\interactionSummary}{
\textbf{\interactionA{}:}
Text-selection + Code Tools menu or right-click Code Tools menu.\\
\textbf{\interactionB{}:}
Box-selection + pop-up Code Tools menu.
}

\interactionSummary{}

\newcommand{\sidebysidetaskquestions}[1]{

Which interaction worked better for the #1 task?

\begin{itemize}
  \item \interactionA{} worked much better
  \item \interactionA{} worked a little better
  \item They are about the same
  \item \interactionB{} worked a little better
  \item \interactionB{} worked much better
\end{itemize}

}
\sidebysidetaskquestions{One Rectangle}

\sidebysidetaskquestions{Two Circles}

\sidebysidetaskquestions{Three Rectangles}

\sidebysidetaskquestions{Target Icon}

Explain your answers to the four previous questions.
When did \interactionA{} work better and why?
When did \interactionB{} work better and why?

\clearpage

\subsection*{Questions about Final Two Tasks}

For the last two tasks, you were allowed to use both
\interactionA{} and \interactionB{}. The following
questions ask about this combination of features.
\\

\newcommand{\likertBox}[1]{\framebox[0.50in][c]{#1}}
\newcommand{\likertOptions}
  {\likertBox{1}\likertBox{2}\likertBox{3}\likertBox{4}\likertBox{5}}
\newcommand{\susQuestion}[1]{#1 & \likertOptions \\\\}

\begin{tabular}{p{3.0in}c}
\renewcommand{\arraystretch}{3.0}
& \textbf{Strongly} \hfill \textbf{Strongly} \\
& \textbf{Disagree} \hfill \textbf{Agree} \\[20pt]
\susQuestion{1. I think that I would like to use this system frequently.}
\susQuestion{2. I found the system unnecessarily complex.} \\
\susQuestion{3. I thought the system was easy to use.} \\
\susQuestion{4. I think that I would need the support of a technical person to be able to use this system.}
\susQuestion{5. I found the various functions in this system were well integrated.}
\susQuestion{6. I thought there was too much inconsistency in this system.}
\susQuestion{7. I would imagine that most people would learn to use this system very quickly.}
\susQuestion{8. I found the system very cumbersome to use.}
\susQuestion{9. I felt very confident using the system.} \\
\susQuestion{10. I needed to learn a lot of things before I could get going with this system.}
\end{tabular}


\newcommand{\fullmodequestion}[1]{

Did the Code Tools (either with Text-Select Mode
or Box-Select Mode) work well for the {#1} task?
If so, how? If not, why not?

}

\fullmodequestion{Four Squares}

\fullmodequestion{Lambda Icon}


\subsection*{Additional Questions}

What computer did you use?

\begin{itemize}
  \item My own personal laptop
  \item The laptop provided by the user study administrator
\end{itemize}

What improvements or new features would make Code
Tools in Sketch-n-Sketch better?


Are there any other comments about Code Tools in
Sketch-n-Sketch that you would like to share?

Are there other languages, application domains, or
settings where you would like to see the Code Tools
features?

\subsection{Additional Discussion}

\parahead{Within-Subjects Experimental Design}

One of our experiments measured which mode participants
preferred when given the ability to mix modes. This was
possible only because of the within-subjects design.
Our within-subjects design also enabled us to control for
each subject's measured skill level (via the random effect
for each participant in the mixed model). A between-subjects
design would require more participants to avoid an imbalance
of participant skill between the treatment groups.
On the other hand, a between-subjects design would be
simpler to interpret and would mitigate concerns about
learning effects between modes. Instead, we relied on the
mixed model to control for learning effects.

\parahead{System Usability Results}

In \autoref{sec:user-study-results}, we discussed the responses to
all survey questions except one (due to space constraints).
Our survey asked users to rate Combined Mode using the System
Usability Scale~\cite{SUS}. The score (mean: 63.9) was in the second
quartile (describable as between ``OK'' and ``Good'') compared to a corpus of
SUS evaluations~\cite{Bangor2009}. There was moderate correlation between
completion rate and SUS score (Pearson's r=0.54); the score among participants
who completed all tasks (extrapolated mean: 75.0) was in the third
quartile (describable as between ``Good'' and ``Excellent'').

 \clearpage

\end{document}